\documentclass[preprint]{emulateapj}
\usepackage{apjfonts}

\newcommand{\galex}{{\it GALEX}}   
\newcommand{\ha}{{H$\alpha$}}
\newcommand{\hb}{{H$\beta$}}
\newcommand{\vmax}{{$V_{\rm max}$}}

\slugcomment{2007 Apr 20 --- ApJS Special \galex\ Issue}

\shorttitle{UV SFRs IN THE LOCAL UNIVERSE}
\shortauthors{SALIM ET AL.}

\begin{document}

\title{UV Star Formation Rates in the Local Universe} 
\author{Samir Salim\altaffilmark{1,*}, 
R.\ Michael Rich\altaffilmark{1}, 
St\'ephane Charlot\altaffilmark{2}, 
Jarle Brinchmann\altaffilmark{3}, 
Benjamin D.\ Johnson\altaffilmark{4},
David Schiminovich\altaffilmark{4},
Mark Seibert\altaffilmark{5}, 
Ryan Mallery\altaffilmark{1},
Timothy M.\ Heckman\altaffilmark{6},
Karl Forster\altaffilmark{7},
Peter G.\ Friedman\altaffilmark{7},
D.\ Christopher Martin\altaffilmark{7},
Patrick Morrissey\altaffilmark{7},
Susan G.\ Neff\altaffilmark{8},
Todd Small\altaffilmark{7},
Ted K.\ Wyder\altaffilmark{7},
Luciana Bianchi\altaffilmark{9},
Jose Donas\altaffilmark{10},
Young-Wook Lee\altaffilmark{11},
Barry F.\ Madore\altaffilmark{5},
Bruno Milliard\altaffilmark{10},
Alex S.\ Szalay\altaffilmark{6},
Barry Y.\ Welsh\altaffilmark{12}, 
Sukyoung K.\ Yi \altaffilmark{11}}

\altaffiltext{1}{Department of Physics and Astronomy, University of
California, Los Angeles, CA 90095 \\
$^{\star}$ Current address: NOAO, 950 North Cherry Ave., Tucson, AZ 85719,
samir@noao.edu} 
\altaffiltext{2}{Institut d'Astrophysique de Paris, CNRS, 98 bis
  boulevard Arago, F-75014 Paris, France} 
\altaffiltext{3}{Centro de Astrof\'{i}sica da Universidade do Porto, 
 Rua das Estrelas 4150-762 Porto, Portugal} 
\altaffiltext{4}{Department of Astronomy, Columbia University, 
  New York, NY 10027}
\altaffiltext{5}{Observatories of the Carnegie Institution of
  Washington, 813 Santa Barbara St., Pasadena, CA 91101}
\altaffiltext{6}{Department of Physics and Astronomy, The Johns Hopkins
  University, Homewood Campus, Baltimore, MD 21218}
\altaffiltext{7}{California Institute of Technology, MC 405-47, 1200
  East California Boulevard, Pasadena, CA 91125}
\altaffiltext{8}{Laboratory for Astronomy and Solar Physics, NASA
  Goddard Space Flight Center, Greenbelt, MD 20771}
\altaffiltext{9}{Center for Astrophysical Sciences, The Johns Hopkins
  University, 3400 N.\ Charles St., Baltimore, MD 21218}
\altaffiltext{10}{Laboratoire d'Astrophysique de Marseille, BP 8, Traverse
  du Siphon, 13376 Marseille Cedex 12, France}
\altaffiltext{11}{Center for Space Astrophysics, Yonsei University, Seoul
  120-749, Korea}
\altaffiltext{12}{Space Sciences Laboratory, University of California at
  Berkeley, 601 Campbell Hall, Berkeley, CA 94720}

\begin{abstract}

We measure star formation rates (SFRs) of $\approx 50,000$
optically-selected galaxies in the local universe ($z\approx0.1$),
spanning a range from gas-rich dwarfs to massive ellipticals. We
obtain dust-corrected SFRs by fitting the \galex\ (ultraviolet) and
SDSS (optical) photometry to a library of population synthesis models
that include dust attenuation. For star-forming galaxies, our UV-based
SFRs compare remarkably well with those derived from SDSS-measured
emission lines (primarily \ha). Systematic deviations from perfect
agreement between these two methods is shown to be due to differences
in the dust attenuation estimates. In contrast to measurements based
on \ha, UV provides reliable SFRs for galaxies with weak or no \ha\
emission, and where \ha\ is contaminated with an emission from an AGN
(1/2 of the sample).  We use full-SED SFRs to calibrate a simple
prescription that uses \galex\ far-UV magnitude and the UV slope to
produce good dust-corrected SFRs for normal star-forming galaxies. The
specific SFR (SFR normalized by stellar mass) is considered as a
function of stellar mass for (1) star-forming galaxies with no AGN,
(2) those hosting an AGN, and for (3) galaxies without \ha\ emission
(the latter two groups forming the bulk of the optical red
sequence). We find that the three have distinct star formation
histories, with AGN lying {\it intermediate} between the star-forming
and the quiescent galaxies. Normal star forming galaxies (without an
AGN) lie on a relatively narrow linear sequence. Remarkably, galaxies
hosting a strong AGN appear to represent the {\it massive
continuation} of this sequence. On the other hand, weak AGN, while
also massive, have lower SFR, sometimes extending to the realm of
quiescent galaxies. We propose an evolutionary sequence for massive
galaxies that smoothly connects normal star-forming galaxies to
quiescent (red sequence) galaxies via strong and weak AGN. We confirm
that some galaxies with no \ha\ emission show signs of star formation
in the UV. We derive a UV-based cosmic star formation density at
$z=0.1$ with significantly smaller total error than previous
measurements.

\end{abstract}


\section{Introduction}

Current studies of galaxies are characterized by two major features:
availability of large samples of objects (galaxy surveys), and the
utilization of the multiwavelength approach. Such studies extend from
the nearby galaxies to those close to the epoch of the formation of
the first galaxies. A rate at which a galaxy forms stars is one of the
more important properties in studying galaxy evolution. The
multiwavelength approach allows us to employ a suite of star formation
(SF) indicators---from X-rays to radio wavelengths. Major effort in
recent years was made to provide reliable calibrations for different
SF estimators and to understand their differences, advantages and
limitations. The most straightforward evaluation is achieved by
comparing two or more star formation indicators for the {\it same} set
of objects. While new galaxy surveys provide large statistical samples
with which one can attempt such studies, the sample of galaxies for
which more than a single star formation indicator can be applied is
not necessarily large. Moreover, various SF indicators are often
applied for different redshift regimes. Global cosmic star formation
history is therefore the result of studies that employ different SF
indicators.

Among the most frequently used star formation indicators are the UV
continuum (usually at $\lambda <2000$ \AA), nebular recombination
lines (primarily \ha, but also [OII]), far-IR dust emission, and the
synchrotron radio continuum at 21 cm \citep{kennicutt_radio}. A
comprehensive review of most of these methods was presented in
\citet{kennicutt}, together with simple formulae for the conversion of
the true flux into a star formation rate (SFR). Recently, the use of
other SF indicators has been explored, such as the X-ray continuum
(e.g., \citealt{xray,xray2}), or the luminosity of PAH features in the
mid-IR \citep{roussel}. Star formation indicators are only as good as
the assumptions that connect a certain observed luminosity to the
actual star formation rate, and therefore all are sensitive to various
systematic uncertainties. One factor has proved more frustrating to
account for than the others---the effect of dust obscuration on the UV
and \ha\ star formation rates. In contrast to these two dust-sensitive
indicators, bolometric IR luminosity and the radio luminosity are
often considered to be ``true'' SF indicators. While this is certainly
an oversimplification, such perception is bolstered by the very tight
correlation of the IR and the radio luminosities, at least as observed
for the normal galaxies in the local universe \citep{dejong}. However,
even if they were perfect, IR or radio methods cannot make UV or \ha\
methods obsolete for many practical reasons.

There have been a number of studies that compared the UV and \ha\ star
formation rates---either as the observed luminosities, or by applying
various schemes to correct for the dust attenuation.\footnote{In this
paper we will use the term ``attenuation'', rather than the more
commonly used term ``extinction'', thus emphasizing the complex
processes of absorption and scattering in a galaxy, rather than the
dimming of light along a single line of sight.} These comparisons were
often carried out with respect to some dust-free SF
indicator. \citet{hopkins} used a sample of several hundred objects
with \ha\ and UV measurements (with the $U$-band photometry serving as
a proxy for the UV), and compared them to the far-IR measurements
compiled by \citet{cram}. They find that in order to reconcile \ha\
and far-IR luminosities one needs to apply a dust attenuation that is
not fixed, but rather depends on the SFR itself. Actually, this
dependence is a consequence of the relationship between the dust
attenuation and the far-IR luminosity found by
\citet{wang}. \citet{hopkins} use an extinction curve to extrapolate
the dust attenuations obtained for \ha\ into the UV regime, but find
that such correction fails to bring UV luminosities to agree with the
far-IR, i.e., the simple application of the same attenuation mechanism to
both the \ha\ and the UV emission does not appear to be 
correct. \citet{bell_ken} used actual UV observations of 50 nearby
galaxies in two ultraviolet bands obtained with the Ultraviolet
Imaging Telescope (UIT), and compared them to \ha\ luminosities from
the narrow-band imaging. They determine UV
attenuation for 13 galaxies using the correlation with the UV slope
\citep{calzetti94}, and measure \ha\ attenuation for 21 objects using
the thermal radio continuum fluxes. They find that both can reach high
levels even for normal galaxies ($\sim 4$ mag for \ha, and $\geq 5$
mag for far-UV), and lend support to previous notions that the
attenuation increases with SFR. \citet{sullivan1,sullivan2} confirmed
that a better agreement between the UV observations from the FOCA
balloon mission \citep{milliard}, and the fiber \ha\ spectra is
achieved when attenuation corrections are taken to be
luminosity-dependent. They also suggest that one perhaps cannot use
simple extinction curve scalings to convert \ha\ attenuations into UV
attenuations. The breakthrough in resolving this problem came with the
introduction of the two-component dust attenuation model of
\citet{cf00}. This model was indeed motivated by the need to produce a
consistent model for dust attenuation affecting \ha\ and UV continuum
photons. It postulates the existence of short-lived (10 Myr) giant
molecular clouds that affect photons producing the
\ha\ line. On the other hand, the attenuation of the UV continuum,
having timescales longer than the lifetime of giant molecular clouds,
is predominantly produced by the diffuse ISM (after the molecular
clouds have dispersed), at levels that are typically 3 times lower
(for a given wavelength) than those in the molecular clouds. In this
paper we will be using the \citet{cf00} model, thus testing it for the 
first time on a large scale.

In addition to systematic trends, we should mention that some previous
studies were finding that the measurement errors are smaller than the
observed scatter. \citet{sullivan1,sullivan2,paramo04} offer the
explanation for this scatter as arising from the differing timescales
over which \ha\ and UV SFRs are sensitive, so that UV could in some
cases (especially in low-mass galaxies) detect a short starburst that
is no longer observable in \ha.

Sometimes, the systematics are present in the observations
themselves. \citet{rosa-gonzales} showed that even the relatively
reliable determination of \ha\ attenuation can be affected by the
systematics when not correcting for the underlying stellar absorption
in Balmer emission lines. Systematics can also arise with galaxy
samples selected at different wavelengths \citep{buat02}. 

The main obstacle in obtaining UV measurements for a large number of
galaxies in the local universe is that they need to be made outside of
Earth's atmosphere. For this reason in 2003 NASA launched the {\it
Galaxy Evolution Explorer} (\galex, \citealt{chris_galex}). \galex\ is
currently conducting the first ever survey of the UV sky. The imaging
(in two UV bands) is executed in several modes---from a shallow
all-sky survey, to the ultra-deep fields. In this paper we are using
measurements obtained in the medium-deep survey that is designed to
image regions of the sky covered by the Sloan Digital Sky Survey
(SDSS). Thus we obtain a large sample of galaxies with both UV and
optical photometry, as well as spectroscopic redshifts. \galex\ and
SDSS data, and the resulting sample, are described in \S\S
\ref{sec:data} and \ref{sec:sample}. Dust-corrected star formation
rates (and some other physical properties such as the stellar mass)
are obtained by comparing the observed colors to stellar population
synthesis models to which the dust attenuation has been applied (\S
\ref{sec:sed}). While we use full UV to optical SED, the SFRs are
essentially constrained by the UV \citep{salim}. The large size of our
sample permits a robust statistical analysis. We compare our UV-based
SFRs to the results of the major study of \citet{b04}, who use SDSS
spectra to derive \ha-based SFRs for $\sim 10^5$ galaxies (\S
\ref{sec:sfr}). They employ an intricate scheme to correct for the
fiber aperture effects. Thus, our study also serves as a check on the
reliability of their methodology. Finally, we discuss star formation
histories of different classes of galaxies (\S \ref{sec:local}) and
derive a UV-based estimate of the global star formation density at
$z=0.1$ (\S \ref{sec:sfrd}).

\section{Data} \label{sec:data}

\subsection{\galex\ data}

Technical aspects of \galex\ telescope, detectors and data products
are presented in \citet{morrissey,morrissey2}. Here we give a summary
of relevant information. \galex\ surveys the sky in either the imaging
or the grism spectroscopy mode. It simultaneously produces a far-UV
(FUV) and a near-UV (NUV) image having a circular field of view of
$1.25 \degr$ diameter. FUV and NUV filters are broadband, with
effective wavelengths of $1528$ \AA\ and $2271$ \AA, respectively. We
will denote the magnitudes measured in these photometric bands as {\it
FUV} and {\it NUV}. A single field of view imaged by \galex\ is called
a {\it tile}. A tile can consist of one or more {\it visits}, i.e.,
individual exposures. \galex\ surveys the sky in several imaging
modes, which differ in the exposure time per tile. In this study we
use the Medium Imaging Survey (MIS), which is designed to maximize the
coverage of the sky that is included in the SDSS. Typical exposure
times in MIS are 1500 s, yielding limiting magnitudes of
$FUV=NUV=22.7$ mag (AB system is used throughout). \galex\ cannot
point in the vicinity of bright sources (usually stars), which
inevitably produces some gaps in the coverage.

\galex\ data used in this paper come from the MIS portion of the
Internal data release 1.1 (IR1.1), which is an expanded version of the
first public GALEX release (GR1) of MIS (in both IR1.1 and GR1 the
same pipeline, version 4, is used to reduce the data and produce
source catalogs).  Dataset consists of 705 \galex\ tiles. Because of
the anomaly with the FUV detector, 98 tiles lack FUV images. The 705
tiles cover 797 sq.deg.\ of the sky. Each \galex\ field is restricted
to $0.6\degr$ radius, close to the maximum field of
view.\footnote{Many other studies with \galex\ restrict analysis to
0.5 or $0.55\degr$ radius field of view. This is because artifacts are
more common near the detector edge, and the PSF becomes
distorted. Since in this study we match SDSS objects to \galex\
sources, the chances of a match with an artifact are small. Also,
while the PSF at the edges is distorted, the total \galex\ flux is not
affected (see also \S \ref{ssec:qual}).}  Because of the overlap
between \galex\ tiles, the total unique area is 741 sq.deg. Source
catalog for each tile was produced using the SExtractor software
\citep{sextractor}.

In our study we use FUV and NUV fluxes measured in Kron elliptical
apertures. We recalculate flux errors because they were incorrect in
the pipeline version 4 reductions (thus also affecting GR1, cf.\
\citealt{trammell}). Kron magnitudes should represent a good measure
of a galaxy's total flux. If both the FUV and the NUV detections are
present, we use the aperture (size, shape and position) derived from
the NUV image to measure the FUV light. We go from NUV to FUV because
we are generally more sensitive in the NUV. We add zero-point
calibration errors of 0.052 and 0.026 magnitudes to $FUV$ and $NUV$
respectively to account for systematic inaccuracies. The calibration
errors were estimated by analyzing the repeat imaging of a calibration
stellar object, and confirmed by a large number of repeat observations
of the same field \citep{morrissey2}. If a UV detection is present in
only one band, we still measure the formal flux and its error in the
other band (using the aperture defined in the detected band). In some
cases, the optical source is not detected at all by \galex, in which
case we note the sky background at the position of the SDSS source and
compute the flux error. In this calculation we use galaxy sizes that
have been derived (using a calibration based on objects detected by
\galex) from SDSS Petrosian radii in $r$-band. Finally, in some cases
the FUV cannot be used because the FUV image does not exist (15\% of
all galaxies).

\subsection {SDSS data}

In addition to the ultraviolet data from \galex, we use optical data
from SDSS Data Release 4 (DR4, \citealt{sdssdr4}). SDSS is providing
five-band broadband photometry ($ugriz$ bands), and the spectroscopic
follow-up of most galaxies with $r\leq 17.77$ (main galaxy
spectroscopic survey, \citealt{strauss}). In addition to the official
data products from the SDSS collaboration, we use value-added galaxy
catalogs produced by MPA/JHU SDSS team. These catalogs include
reprocessed SDSS spectroscopic data, and some derived galaxy physical
parameters that are based on the spectroscopic data.

SDSS catalog lists magnitudes measured in several different ways. We
use {\sc MODELMAG} magnitudes, which are the measurements of choice
for relative fluxes (i.e., provide stable colors), while still
capturing most of the total light. We apply slight adjustments
($-0.04$ and 0.02 mag) to $u$ and $z$ magnitudes respectively to bring
them closer to the actual AB system \citep{sdssdr2}. We convert SDSS
magnitudes and errors into fluxes using the transformations given in
\citet{scranton}. In addition to catalog photometric errors, we add
0.01 mag of uncorrelated calibration error to each of the bands (see
\S \ref{ssec:qual}), and, specifically for the $u$ band, we add an
additional color-dependent error due to the red leak ($\sigma_{u, {\rm
RL}} = 0.0865(r-i)^2+0.0679(r-i)$, based on \citealt{sdssdr2}). In
rare cases an SDSS magnitude would be missing and such band has to be
excluded from the analysis. We further exclude from analysis
individual SDSS magnitudes that are nominally fainter than $ugriz=25$.
These are invariably spurious, regardless of the listed photometry
error. Finally, we exclude magnitudes with errors larger than
$10\times$ the typical error in that band (we find typical errors to
be 0.07, 0.009, 0.007, 0.007 and 0.017 mag for $u$, $g$, $r$, $i$, and
$z$ bands respectively). While large, we find these errors to be
significantly underestimated. To summarize, except in cases listed
above when we exclude certain individual flux points, our input data
consists of 7-band photometry and spectroscopic redshifts from SDSS.

\subsection {\galex/SDSS matched catalog}

Of 741 sq.deg.\ of \galex\ unique imaging, 645 sq.deg.\ overlaps with
SDSS DR4 spectroscopic area, thus defining the solid angle of our
sample.\footnote{Matching with the current SDSS data release DR5
(which uses the same pipeline as DR4) would not significantly increase
the overlap, since current \galex\ pointings mostly follow the
footprint of SDSS DR1 and DR2.}  We estimate this area by counting the
SDSS galaxies that: (1) have spectra, (2) have a dereddened magnitude
$14.5<r_{\rm Petro}<17.5$ (the faint end is taken to be comfortably
brighter than the spatially variable spectroscopic limit) and (3) lie
at redshifts $0.005<z<0.22$. We count such galaxies both in our
survey, and in the full DR4, whose spectroscopic area is known. The
ratio of the two counts gives us the size of our survey area. This
estimate should be good to within 1 sq.deg. There are 67,883 objects
from the SDSS DR4 spectroscopic survey (not restricted to the main
galaxy survey) that lie in this area, and are spectroscopically
classified as galaxies. For each of these objects we search for a
match in the \galex\ source catalog (which already combines FUV and
NUV detections) within $4''$. Our analysis of SDSS point sources with
a match in \galex\ indicates that \galex\ positions have a random
error of $0\farcs 8$ in either R.A.\ or declination (becoming somewhat
larger at the edges of the field). In addition, there are overall
tile-to-tile offsets between \galex\ and SDSS coordinate systems of
several tenths of an arcsecond. In any case, astrometric uncertainties
are significantly smaller than our matching radius (see also
\citealt{trammell}). A genuine match can be missed if the centroid of
the optical light is offset by more than $4''$ compared to the
centroid of UV light. We expect such cases to be quite rare, since at
the mean redshift of the sample ($z= 0.104$) this offset would
translate into a 7 kpc separation. A problem with matching in general
is that what is considered to be a single object in one catalog can be
resolved into multiple objects in another catalog, whether they are
indeed separate objects (blending), or actually belong to the same
system (shredding). This problem is more pronounced when combining
catalogs with different resolutions, as is the case here (4 to $5''$
for \galex\ vs.\ $1\farcs4$ for SDSS). However, in our particular
sample this problem is not severe. SDSS galaxies with spectra are
relatively bright objects, and if more than one \galex\ object is
found within the search radius our procedure was to take the one that
is positionally the closest. Since we are dealing with high-latitude
fields where the density of foreground stars is not that high, this
simple matching procedure produces reliable results. Besides, we do
have mechanisms of identifying the majority of incorrect matches at
the later stage, when we perform an SED fitting to the combined
\galex/SDSS photometry (see \S \ref{ssec:qual}). Since we combine
photometry from two different catalogs there is a concern of
photometric zero-point mismatch. We explore such a possibility in \S
\ref{ssec:comp_sf}. In our final matched catalog we eliminate
duplicate \galex\ observations (stemming from overlaps or repeat
observations) by keeping those that lies closer to the center of
\galex\ field of view.

\section {The sample} \label{sec:sample}

\subsection{Sample selection}

We initially define our sample by applying magnitude and redshift cuts
to galaxies with SDSS spectroscopy (note that objects
spectroscopically classified as QSOs are excluded). We require the
dereddened Petrosian magnitude to be in the $14.5<r_{\rm
Petro}\leq17.77$ range, where the faint limit is the nominal limit of
the SDSS main galaxy spectroscopy survey (see also \S
\ref{ssec:kcor}), while the bright limit is chosen to avoid objects
with saturated SDSS photometry. We require redshifts to lie within
$0.005<z\leq 0.22$ range. The lower limit is chosen to eliminate
galaxies that may deviate from the Hubble flow, i.e., whose redshift
distances could be unreliable. Redshift distribution in SDSS main
galaxy sample peaks sharply around $z=0.1$ with few galaxies beyond
our upper redshift limit. These magnitude and redshift limits are
identical to those used in \citet{b04}, to which we will be comparing
many of our results. We will refer to the above redshift range as
representing the ``local universe''.

There are 49,346 galaxies that meet the conditions defining our
initial sample. Good quality SED fitting (see \S \ref{ssec:fit}) is
obtained for 98\% of them (48,295 galaxies).  In the rest of the paper
we will use only this {\it latter} sample and refer to it as ``all''
galaxies. Note that since we retain objects regardless of whether they
were detected by \galex\ (as long as they fall within the area covered
by \galex), our sample is only {\it optically} ($r$-band) selected.

Throughout the paper we will use $\Omega_m=0.3$, $\Omega_\Lambda=0.7$,
$H_0= 70\, {\rm km\, s^{-1}\, Mpc^{-1}}$ (i.e., $h_{70}=1$) cosmology.

\subsection{Emission line diagnostics and 
sample classification} \label{ssec:bpt}

This work focuses on the physical properties of galaxies, such as
their star formation rates (SFRs) and stellar masses. Before we start
analyzing the sample based on these {\it derived} properties, we would
like to characterize it in terms of its {\it observable} quantities.

Throughout this paper we will rely on optical emission lines to
classify galaxies in our sample. This classification, based on the BPT diagram
\citep*{bpt}, plots the flux ratio of [OIII]$\lambda$5007\AA\ and \hb\
lines against the flux ratio of [NII]$\lambda$6584\AA\ and \ha\
lines. In this paper we fully adopt the BPT classification of
\citet{b04} (hereafter B04), which is based on emission-line
strengths corrected for the underlying stellar absorption (see their
Fig.\ 1). The position of a galaxy in the BPT diagram indicates the
nature of its ionizing source. The majority of galaxies in the diagram fall
within one of the two branches. One branch lies mostly above the
maximum line ratios expected from star formation. These
galaxies must have some ionizing source in addition to young stars. Most
researchers attribute this emission to an active galactic nucleus
(AGN). Specifically, the emission is associated with a narrow-line AGN
(a LINER or a Seyfert 2), also known as Type 2 AGN
\citep{kauff_agn}. Note that any Type 1 AGN spectrum (QSO or Seyfert
1) would have broad lines, and would be classified by the SDSS
pipeline as a QSO and thus not included in this sample. Following
\citet{kauff_agn}, B04 classify galaxies that lie on the lower portion
of the AGN branch, but with line ratios allowed by star formation, as
``composite'' (star forming/AGN) galaxies. For galaxies
that lie on the star forming branch we expect very little
emission line flux to come from an AGN \citep{kauff_agn}. 
B04 required a S/N ratio in
all four lines to be $>3$ in order to apply the BPT diagram
classification. However, in some cases a secure AGN classification is
possible even when only [NII]6584 and \ha\ have S/N$>3$, i.e., in
cases when this ratio is larger then the one allowed by SF. B04 calls
this class a ``low-S/N AGN (LINER)''. Following B04, we will study
this class together with the AGN. B04 also introduce the category of
``low-S/N SF'' galaxies. These are the galaxies that do not fall in
any of the previous categories because their lines have low S/N, yet
they have an \ha\ detection with S/N$>2$. While their lines are not
strong enough for secure placement on the BPT diagram, if we
nevertheless do so, we find that many object lie in the high-mass end
part of the SF sequence, as well as in the lower portion of the AGN
branch. Therefore, these objects represent a heterogeneous
class. Finally, there are galaxies without detectable lines, thus
precluding the classification in the above scheme. B04 call this group
``unclassifiable''. We will call them ``No \ha'' class. We find that
in $\sim 2\%$ of galaxies in this class the \ha\ non-detection is due
to some artifact in the spectrum or line-extraction pipeline. We
exclude these galaxies from this class (but not from the whole
sample). The fraction of galaxies in our sample belonging to different
classes is as follows: Star forming (SF) -- 27\%, Low S/N SF -- 19\%,
Composite (SF/AGN) -- 8\%, AGN -- 12\% and No \ha\ -- 33\%.

\subsection{UV-to-optical color-magnitude diagram} \label{ssec:cmd}

\begin{figure*}
\epsscale{1.2}
\plotone{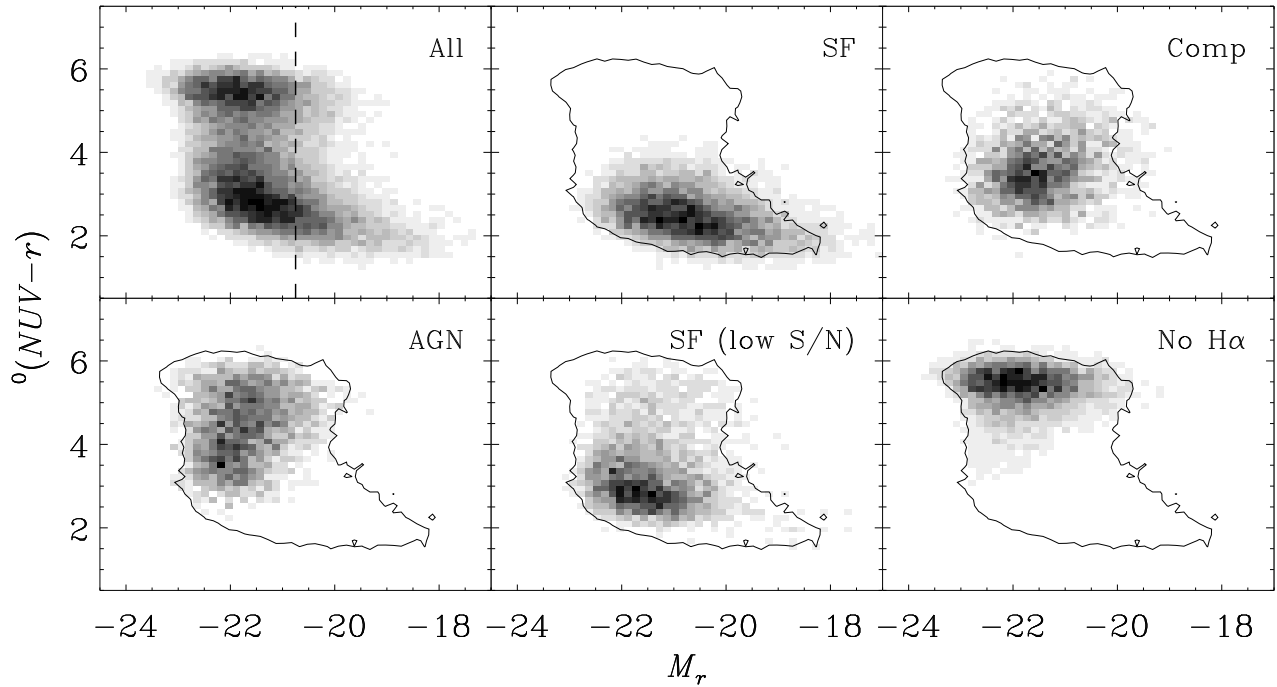}
\caption{UV to optical color-magnitude diagrams (CMDs). Upper left
  panel shows a greyscale scatter plot of all galaxies in our sample
  with a near-UV detection. (The shade of gray is directly correlated
  to the number of points contained in a given ``pixel''.) Note the
  pronounced bimodality of the blue and the red sequences, and their
  large separation. Dashed line represents the completeness limit at
  the mean redshift of our sample. The remaining panels show CMDs of
  different galaxy classes (SF -- star forming, Comp --
  star-forming/AGN composite, AGN -- Type 2 AGN, SF (low S/N) --
  star-forming with weak \ha, and No \ha), as determined from the
  position in the BPT diagram. The outer contour encompasses 90\% of
  the entire sample is plotted for reference. While SF galaxies mostly
  lie in the blue sequence, and those with no \ha\ in the red, most
  galaxies in between the two sequences are AGN or AGN/SF
  composites. Absolute magnitude is given in $z=0$ $r$-band, and the
  color is K-corrected to $z=0$ rest-frame, as indicated by
  superscript 0.}
\label{fig:cmd}
\end{figure*}

Color-magnitude diagram (CMD) is a powerful tool in assessing the
basic properties of a sample of galaxies. Historically, the study of
CMDs was preceded by the studies of a relationship between color and
morphology. Basic morphological segregation into disk-like spiral
galaxies and spheroidal elliptical galaxies was established by
\citet{hubble1}. Afterward (e.g., \citealt{hubble2}) it was realized
that spiral (late type) galaxies have bluer colors than the
ellipticals (early types). Optical CMDs were constructed for cluster
early type galaxies (e.g., \citealt{sandage}), where they were found
to form a narrow sequence (the so called red sequence). Field spiral
galaxies also displayed the color-magnitude relationship
\citep{chester}, albeit with larger scatter. The bimodal nature of the
field galaxy CMD became much more apparent with recent large scale
surveys, in particular SDSS \citep{baldry}. However, unlike the
traditional optical CMDs, a CMD in which the color is constructed from
an {\it ultraviolet} and an optical magnitude has a particular
diagnostic power. By virtue of contrasting the current (or recent)
star formation as indicated by the UV light (modulo attenuation) to
the total past star formation as indicated by the optical light, the
UV to optical color is a good proxy of a galaxy's SF history (e.g.,
\citealt{salim}). On the other axis, one plots absolute optical
magnitude, which is related (modulo variations in the optical
mass-to-light ratio) to galaxy's current stellar mass. We present the
observed CMD of our sample in Figure \ref{fig:cmd}. In this and many
subsequent figures, the individual data points have been converted
into a greyscale density scatter plot, in such a way that the shade of
gray is proportional to the number of objects occupying a given
``pixel''. Such representation is desirable when dealing with large
samples where it is easy to saturate a traditional scatter
plot. Unless noted otherwise, the figures are constructed from raw,
unweighted counts.  We will include volume corrections later, when
appropriate. Both the color and the absolute magnitude have been
K-corrected to $z=0$ rest-frame bands (\S \ref{ssec:kcor}). The upper
left panel shows the CMD of all galaxies with an NUV detection (85\%
of the total sample; see Table \ref{tab:errors} for a breakdown of UV
detection rates per class). A striking feature of a UV to optical CMD
is the pronounced bimodality of blue and red galaxies. Blue galaxies
form a well-defined sequence extending to faint luminosities. The red
sequence is somewhat more narrow than the blue sequence (note that
this becomes evident only after K corrections have been applied), and
extends to intrinsically more luminous galaxies than the blue
sequence. This is related to well-known fact that the most massive
elliptical galaxies are more massive than the most massive spirals
\citep{holmberg}. The two sequences are separated by $\sim 3$ mag. If
the two sequences are modeled as gaussians, one finds that there is an
excess of galaxies in the gap \citep{wyder}. This is not the case in
classical optical CMDs \citep{baldry}. We will refer to the gap region
and its population as the ``green valley'' \citep{chris2}. A detailed
quantitative study of the \galex\ UV to optical CMD is presented in
\citet{wyder}.

\begin{deluxetable}{lrrrr}
\tablewidth{0pt}
\tablecaption{Average errors of ``UV'' and ``\ha''-based 
  star formation rates\tablenotemark{a}}
\tablehead{ Class & Number & UV detected\tablenotemark{b}  &
``UV'' & ``H$\alpha$'' \\
 & & &  $\langle \sigma({\rm log~SFR}) \rangle$ & 
 $\langle \sigma({\rm log~SFR}) \rangle $ }
\startdata
All\tablenotemark{c}  &  48295  &  86\%  &  0.38  &  0.43 \\
SF              &  12901  &  99\%  &  0.20  &  0.29 \\
SF (low S/N)    &   9060  &  93\%  &  0.30  &  0.39 \\
Comp            &   3966  &  96\%  &  0.28  &  0.40 \\
AGN             &   5827  &  90\%  &  0.41  &  0.49 \\
No \ha          &  16159  &  68\%  &  0.60  &  0.54 \\
\enddata

\tablenotetext{a}{UV-based SFRs are averaged over 100 Myr.}
\tablenotetext{b}{Either FUV or NUV detection. Of objects detected in
  FUV, 98\% are also detected in NUV.}
\tablenotetext{c}{Includes 382 objects for which classification was
  not possible, see \S \ref{ssec:bpt}.}
\label{tab:errors}
\end{deluxetable}

In the subsequent panels in Figure \ref{fig:cmd} we display the CMDs
of various classes of galaxies as defined in \S\ref{ssec:bpt}. Each
panel has the greyscale normalized to the number of galaxies in the
given class. For better reference with respect to the full sample, in
each subsequent panel we repeat the contour containing 90\% of {\it
all} galaxies. Not surprisingly, the star-forming galaxies (SF) occupy
the blue portion of the CMD. Part of the width of the sequence is due
to the intrinsic dust attenuation. The CMD of composite galaxies
(showing both signatures of SF and AGN) are shown in the upper right
panel. Their $NUV-r$ colors are offset to the red compared to those of
the ``pure'' SF galaxies.  Galaxies with narrow line AGN (lower left
panel) occupy the regions of the red sequence {\it and} of the green
valley (\citealt{chris2,kauff_galex}). Moreover, most galaxies with
intermediate colors are either AGN or composites. Star formation
histories of AGN will be discussed more extensively in \S
\ref{ssec:ssfr_agn}. Next, we have low-S/N SF galaxies (middle lower
panel), which indeed mostly lie on the blue, star-forming sequence
(preferentially its more luminous part). However, there is a tail of
red galaxies which are probably contaminated by AGN as observed in \S
\ref{ssec:bpt}. Finally, we have galaxies with no detectable \ha\
(lower right panel). As expected, these galaxies form the bulk of the
red sequence. However, there is a tail of galaxies of this class
extending towards the blue colors. Since the spectral classification
used here is nuclear (central $3''$) it is possible that some of these
galaxies are classified as not having \ha\ because of the dominant
bulge, while the relatively faint star forming disk is giving a galaxy
an overall blue color. However, as we discuss in \S
\ref{ssec:ssfr_noha}, this is not the case for most of them. The CMDs
presented in Figure \ref{fig:cmd} require a detection in NUV. The
remaining 15\% of our sample with no NUV detection falls mostly in the
red sequence (as evidenced from their $u-r$ colors). These galaxies
are too faint to be detected in the UV.

\section{Obtaining galaxy properties with SED fitting} \label{sec:sed}

\subsection{SEDs of model galaxies} \label{ssec:models}

Spectral energy distribution (SED) fitting is becoming a widely used
technique for deriving galaxy properties. It was pioneered in works of
\citet{searle} and \citet{lar_tin}. To first order it consists of
comparing the observed SED to a set of model or template SEDs, and
searching for the best match. It is assumed that the physical
properties of model or template galaxies are known, and that one can
use this knowledge to deduce the properties of an observed galaxy.

In this study, we compare the observed SED to a large number of model
SEDs constructed from \citet{bc03} population synthesis code. To
construct the models we first choose input model properties by
randomly selecting values from prior distributions defined in the
following way. Formation of a galaxy is uniformly distributed between
0.1 Gyr and the age of the universe. Global metallicity is uniformly
distributed between 0.1 and 2 $Z_{\odot}$. Star formation histories
are not single stellar populations, but the combination of an
exponentially declining continuous star formation,
SFR$(t)\propto\exp(-\gamma t)$, with $0\leq\gamma\leq1\,$Gyr$^{-1}$
uniformly distributed over that range, and the random {\it starbursts}
superimposed on the continuous SF.  Bursts are parametrized to have a
duration uniformly distributed in the 30--300 Myr range, with a
strength such that the mass produced in a burst is between 0.03 and 4
times the mass produced in the continuous SF over the present lifetime
of a galaxy (distributed uniformly in log). Such parametrization
conforms with most observational studies. Finally, bursts are randomly
produced so that the probability that a given galaxy has undergone at
least one burst over a 2 Gyr period is 50\%. The above parameters
define how the population synthesis for each model will be carried
out. Next, each of the above model SEDs is subjected to dust
attenuation parametrized according to the two-component model of
\citet{cf00}. For the $V$-band optical depth we randomly chose a value
from a distribution $0<\tau_{V}<6$ (it peaks at $\tau_V=1.2$ and has
most values in the $0<\tau_{V}<2$ range). The prior distribution for
$\tau_V$ is empirical, and comes from Balmer decrements in SDSS
spectra. The choice of $\tau_V$ prior distribution will be discussed
in \S \ref{ssec:comp_sf}. Fraction of the optical depth that affects
stellar populations older than 10 Myr (i.e., most of the UV continuum
flux) is given by factor $\mu$, with values ranging from 0.1 to 1,
peaking around $\mu=0.3$. Altogether, we produce 100,000 model spectra
spanning a range of SF histories, metallicities and dust attenuations.
Note that since we pick input parameters randomly, we do not call our
set of models a grid, which would suggests a set of points with equal
spacing in parameter space.

Finally, we convolve the resulting model spectra with \galex\ and SDSS
bandpasses at five redshifts equally spaced in the [0.05, 0.25]
interval, producing the libraries of model broad-band photometry. In
each library we keep only models that have an age smaller than the age
of the universe at that redshift. This effectively reduces the number
of models from 95,000 at $z=0.05$ to 78,000 at $z=0.25$. We also add
effects of the intervening intergalactic extinction according to
\citet{madau}. Our final libraries list model photometry as well as a
number of galaxy parameters associated to that model (such as the SFR
averaged over several timescales, stellar masses, dust attenuation
parameters, etc.)

\subsection{SED fitting} \label{ssec:fit}

In our study, the observed SED is constructed from \galex\ and SDSS
broadband photometric fluxes (the broad-band SED). For an observed
galaxy at some redshift we select the model library with the closest
redshift. We step through a library one model at the time. Model flux
points will have some arbitrary zero point, and in order to see how
well an observed flux compares to the model flux we first need to find
a factor $a$ that minimizes the $\chi^2$ between the observed and the
model points. In other words, for model $i$, we need to minimize the
following expression:

\begin{equation}
\chi^2_i = \sum_X \left(\frac{F_{{\rm obs},X}-a_i F_{{\rm mod}_i,X}}
{\sigma(F_{{\rm obs},X})}\right)^2 \label{eqn:chi2}.
\end{equation}

\noindent Here, the summation is over $X=(FUV,NUV,u,g,r,i,z)$, the
observed flux points are $F_{\rm obs}$, and their errors
$\sigma(F_{\rm obs})$. Flux points of model $i$ are $F_{{\rm
mod}_i}$. By taking the derivative of Equation \ref{eqn:chi2} with
respect to $a$, and equating it with zero, we find the scale factor
$a^*$ that best matches the observed and the model flux points:

\begin{equation}
a^*_i = \frac{\sum_X \frac{F_{{\rm obs},X} F_{{\rm mod}_i,X}}
{\sigma^2(F_{{\rm obs},X})}} {\sum_X \left(\frac{F_{{\rm
mod}_i,X}}{\sigma(F_{{\rm obs},X})}\right)^2}.
\end{equation}

Because we include non-detections that have meaningless fluxes (small
or even negative values), we use the observed fluxes {\it without}
correcting for Galactic reddening, and instead apply the reddening to
model fluxes, which are always positive. This allows us to treat
non-detections like any other flux point with a known flux
error. Reddening corrections for SDSS magnitudes come from SDSS
database, and those for \galex\ from relations in \citet{wyder} (Their
Eq.\ 2 for $NUV$).

The goodness of the fit between an observed galaxy and model $i$ is
then obtained by inserting $a^*_i$ in place of $a_i$ in Equation
\ref{eqn:chi2}. Note that only a single parameter is being fit---the
scale factor $a$. We emphasize that the word ``parameter'' when
fitting is discussed should not be confused with (in principle)
arbitrary number of {\it galaxy} parameters (i.e., properties) that
{\it correspond} to each model SED. Therefore, since we have seven
photometric points, there are six degrees of freedom. For a given
galaxy we evaluate $\chi^2_i$ for {\it each} model as outlined
above. Now, each of these $\chi^2_i$ values will determine the
relative {\it weight} $w_i$ for the given model $i$ (and therefore the
weight of the galaxy parameters associated with that model) as $w_i=
\exp(-\chi^2_i/2)$. We then build a probability distribution function
(PDF) for every galaxy parameter of interest by compounding weights at
the corresponding parameter value. We repeat the procedure for each of
the $\sim 10^5$ models in the given redshift library. We normalize the
final PDFs and note the parameter values corresponding to the 2.5, 16,
50 (median), 84 and 97.5 percentiles of the cumulative PDF. We use the
average of the PDF as our nominal estimate of the parameter
value. Since in the general case the PDF is not symmetric, medians,
averages and modes will differ. For most relevant parameters in this
study the differences between the median and the average are small. We
do not consider using a mode, because due to the discreteness of the
model parameter space, the mode can be stochastically offset from the
bulk of the PDF, and is sensitive to PDF binning. For similar reasons
we do not use a {\it single} model with the best $\chi^2$ (i.e., the
maximum likelihood model) as representative of parameter values, but
we do keep track of the best $\chi^2$ values in order to evaluate the
overall quality of the fitting for a given galaxy (\S
\ref{ssec:qual}). Additionally, for certain parameters we preserve
more detailed information on the shape of the PDF (in the case of the
specific SFR and stellar mass, we actually keep their mutual two
dimensional PDFs). In certain cases we will discuss the errors of the
derived parameters. We use 1/4 of the 2.5-97.5 percentile range as a
proxy for what would have been a $1 \sigma$ error in the gaussian
distribution. We will refer to these quantities as ``formal'' errors.

We prefer the above-described Bayesian approach (e.g.,
\citealt{bayes}) to often-used maximum likelihood, since it allows one
to transparently test the dependence on the prior, i.e., how well the
data constrain certain parameters given that the shape of the
resulting PDF can to some extent depend on the distribution of the
given parameter in the model library (especially when the
observational constraints are poor). Another advantage is that while
the maximum-likelihood model might indeed correspond to a high
probability density, in the case of extensive models such as ours, it
might actually correspond to a negligible probability "mass", and
might therefore lead to misleading parameter estimation (see for
instance the discussion in \citealt{mackay}). Note also that our
technique properly accounts for the degeneracies inherent in some
galaxy properties. Imagine that we have two galaxy parameters whose
values can be picked in such a way as to produce {\it identical} model
SEDs. In such case our fitting will give {\it equal} probabilities to
all of the degenerate models, and the resulting PDFs for each of these
two parameters will therefore reflect the entire range that these
parameters can take (i.e., there will be nothing to constrain the
PDFs). If we then characterize the errors of those parameters using
the width of the PDF, we will obtain very large values, i.e., we will
know that a certain parameter is poorly constrained. This, in effect,
is in contrast to fitting using a {\it single} best-fit model, which
would pick one of the degenerate models as the best-fitting (and the
corresponding parameter value), without giving a researcher an idea
that there is an underlying degeneracy making the obtained value to be
quite uncertain. Finally, we want to emphasize that the widths of the
PDFs (and also their typical values) have a meaning only if the
observational errors have been estimated reasonably correctly, and if
the overall fit (which can be well characterized by the $\chi^2$ of
the best-fitting model) is reasonably good (\S \ref{ssec:qual}).

\subsection{Quality of a fit} \label{ssec:qual}

For our initial sample of 49,346 galaxies we obtain model parameters
in the way described in \S \ref{ssec:fit}. For each galaxy we also
note the $\chi^2$ of the {\it single} best-fitting model, $\chi^2_{\rm
best}$. This value serves as an indicator of the overall quality of
the fitting for a given galaxy. Since we include galaxies close to the
edge of \galex\ field of view where we know that image PSFs get
distorted, we would first like to check if this affects the
photometry, and therefore the quality of SED fitting. Thus we look at
the average $\chi^2_{\rm best}$ as a function of distance from the
\galex\ image center. We find no correlation with the distance or
degradation close to the edge.

In the ideal case, $\chi^2_{\rm best}$ values would form a theoretical
$\chi^2$ distribution. The average of such distribution should equal
the degrees of freedom (in our case 6). Before looking at the $\chi^2$
distribution we notice that $\chi^2_{\rm best}$ values are
systematically higher for blue galaxies compared to the red. We
believe that this difference arises from the presence of emission
lines in blue galaxies that effectively increase the discrepancy
between the observed SED and the model (line-free)
continuum. Therefore, we start by comparing the observed and the
theoretical $\chi^2$ distribution of red galaxies.  Initially, the
observed distribution was shifted towards the values that were too
low. This was indicative of photometry errors being overestimated. We
arrive at the $\chi^2$ distribution that was well-matched to the
theoretical one when we adopt 0.01 mag calibration errors in each of
the 5 SDSS bands.\footnote{The ``official'' SDSS calibration errors
are given as 2\% for $r$ band, and between 2\% and 3\% for each of the
four {\it colors}. However, the individual {\it magnitude} errors
(which figure in Equation \ref{eqn:chi2}) are correlated, which is why
we find them to be smaller than 2\%. Our results stay practically the
same even if errors are increased, but the $\chi^2$ distribution then
deviates from the expected one.} While now the bulk of the red
galaxies follows a $\chi^2$-like distribution, there is also present a
tail of high $\chi^2_{\rm best}$ values, in numbers well above any
expected distribution. One can imagine that the main reason for the
presence of $\chi^2$ outliers is that the data do not reflect the
actual SED for some reason. We will investigate possible causes
momentarily. The high-$\chi^2$ tail of red galaxies contains some 2\%
of objects. We decide to exclude these galaxies from further analysis
(although we do account for them statistically when making any kind of
space density calculation). Assuming that the reasons driving extreme
$\chi^2$ values are color independent, we also clip 2\% of the
blue-galaxy $\chi^2_{\rm best}$ distribution tail. This leaves us with
48,295 galaxies---our final sample.

While removing the outliers is necessary to have results not be biased
by unreliable values, it is instructive to understand what causes some
$\chi^2$ to assume such high values. We thus examine 22 galaxies with
extreme $\chi^2_{\rm best}$ values, distributed evenly in the
$-1<NUV-r<7$ range (thus sampling galaxies with various SF
histories). In 10 cases we find obvious problems with cataloged
photometry. This is most often the case with SDSS (9 objects), where
the visual inspection of broadband SEDs and comparison with spectra
indicates that some flux points are large outliers.  Next, in 7 cases
we are dealing with either close galaxies being blended in \galex\
(4), or with SDSS shredding galaxies into smaller photometric objects
(2). For one object it is unclear whether we have blending or
shredding. Further, in one case an object shows a clear QSO spectrum,
i.e., the continuum is affected by the AGN, yet it is
spectroscopically (mis)classified as a galaxy. Finally, in four cases
it appears that the broad-band SEDs (and spectra) are genuinely
unusual. While their number is too small to affect the present study,
this type of outliers deserves a special scrutiny. It needs to be
explored if there exists any set of model input parameters (SFH, dust
attenuation, metallicity) beyond our already wide range, which can
produce a model SED matching these observations. For our full sample
we also expect on the order of 10 cases where a foreground star of
similar brightness is superimposed with a galaxy, and falls within the
fiber, thus producing a ``composite'' galaxy/star spectrum which would
fail the fitting.

Since our model libraries are calculated at some fixed redshifts,
there is some error associated with the library redshift not perfectly
matching the galaxy redshift. We thus contrast the $\chi^2_{\rm best}$
values of galaxies having the maximum redshift offset (close to 0.025)
to those with no offset. We find a systematic increase in $\chi^2_{\rm
best}$ of 10\%, which is too low to affect the overall $\chi^2$
distribution.

Finally, we want to assess the effects of the ``resolution'' of our
model parameters, i.e, we can ask how extensive should a model library
be that the resolution is not an issue. To that extent, we
artificially degrade our model libraries by running full SED fitting
using every 2nd, every 4th and every 8th model SED. We then compare
average $\chi^2_{\rm best}$ values from each run. Overall, the quality
of the fits are surprisingly stable. The run with every 2nd model
produces only a 4\% increase in $\chi^2_{\rm best}$ values, while the
run using every 16th model is 16\% worse compared to the full
library. This implies that increasing the extent of the library (for
the given range of input parameters) would not bring significant
improvements. This, of course, is not a general conclusion, but rather
states that with the precision of our data the current number of
models is adequate, i.e., having a finer resolution of input model
parameters would create model SEDs that are degenerate from the
observational point of view. In addition to this, the \citet{bc03}
models themselves are not likely to be of such an accuracy to warrant
more extensive model library.

\subsection{K corrections and volume corrections} \label{ssec:kcor}

For deriving the galaxy properties from the SED fitting, we are not
required to know the K corrections. This is because for a galaxy at a
given redshift we compare the observed magnitudes to model magnitudes
with bandpasses that have been shifted to the matching galaxy's
redshift (i.e., the redshift of the model library). Therefore, the ``K
corrections'' are present in the SED fitting implicitly. However, for
some applications, such as constructing a CMD (\S \ref{ssec:cmd}), or
obtaining \vmax\ weights (\S\S \ref{sec:local},\ref{sec:sfrd}), we
need to know the K corrections explicitly. The process of SED modeling
allows us in principle to derive the K corrections alongside galaxy
parameters. It even has the advantage of allowing one to construct a
PDF for any K-corrected magnitude, which thus yields an estimate of a
K correction {\it error}. We find this approach to be especially
useful in K correcting higher redshift ($z\sim1$) samples (Salim et
al., in prep.)  In the current case, given that our redshift range is
covered by only 5 model libraries, we choose to calculate the K
corrections using the publicly available code {\sc KCORRECT
V4\_1\_4}\footnote{Available from
\url{http://cosmo.nyu.edu/blanton/kcorrect}.}  \citep{blanton}. The
code allows \galex\ magnitudes to be used together with SDSS
magnitudes to constrain the SED fit from which K corrections are
derived. We obtain K corrections using the same combined \galex/SDSS
photometry used for the SED fitting, except that $<3\sigma$ UV
detections are not used. We derive K corrections without a priori
assuming an evolution of the luminosity function. We will explicitly
correct for evolution where appropriate.

In cases when we require a volume-corrected sample, we assign a weight
to each galaxy according to its $V_{\rm max}$ value---the volume in
which a galaxy would be visible taking into account redshift and
apparent magnitude limits, and the solid angle of a survey. While our
faint limit ($r=17.77$) is also the nominal limit of the SDSS main
spectroscopic survey, we take into account that the actual
spectroscopic magnitude limit varies from one spectroscopic plate to
another (i.e., that it can be brighter than $r=17.77$). Thus, for each
object we modify the volume by the spectroscopic completeness (usually
around 0.9). We take spectroscopic limits and completeness values from
NYU-VAGC catalog \citep{vagc}. We caution that in order to calculate
$V_{\rm max}$ correctly, it has to be done iteratively, because the K
correction depends on redshift limits that determine $V_{\rm max}$.

\section{Comparison of ``UV'' and ``\ha'' Star Formation Rates}
\label{sec:sfr}

\subsection{``UV'' SFR estimates}

We derive our SFRs from photometric constraints that extend from the
UV to the $z$-band. However, it is the UV luminosity that constrains
the SFR the most. Therefore, in the remainder of the paper we will
refer to SFRs derived from the SED fitting as ``UV'', where quotation
marks stand to remind us that it is not only UV information that
contributed to these SFR estimates. In our model SED libraries we
report ``current'' star formation rates averaged over several time
intervals, most notably over the most recent 10 Myr, 100 Myr and 1
Gyr. By comparing the formal errors of SFR estimates (derived from 95
percentile range of each galaxy's SFR PDF), we find that the
best-constrained SFR for the overall sample is the one averaged over
the last 100 Myr (Table \ref{tab:errors}), which is the timescale for
UV bright stars. For the star-forming class specifically, we find that
the SFRs have slightly smaller formal errors over 1 Gyr timescales
than over 100 Myr.  On the other hand, SFR estimates on timescale of
10 Myr for SF galaxies (a timescale that matches \ha\ SFR) have
drastically larger formal errors, as expected since we do not have
information to constrain them. In the remainder of the paper we will
use SFR estimates averaged over 100 Myr, which represent a good
compromise between the quality and a timescale that is not too long
compared to that of the \ha\ SFRs.

\subsection{``\ha'' SFR estimates} \label{ssec:ha}

B04 have developed a method of deriving SFRs from SDSS spectra that is
primarily based on the intensity of an \ha\ line. This represents the
largest currently available sample of precise \ha-based
SFRs. Obtaining accurate SFRs from SDSS spectra is far from
straightforward. For the purposes of this paper we briefly describe
the procedures employed in B04. The essence of the B04 approach was to
model SDSS spectra by first removing the absorption line spectrum
using a combination of burst models from \citet{bc03}, and then to
model the emission lines using the \citet{cl01} models which combine
the \citet{bc93} galaxy evolution models with the {\sc CLOUDY}
photoionization code \citep{ferland96}. Also, the same two-component
dust attenuation prescription of \citet{cf00} was used by B04 to model
attenuation of the emission lines, and by us to model broad-band
continuum attenuation. While conventionally the attenuation of \ha\
flux is determined by comparing the observed \ha\ to \hb\ ratio
(Balmer decrement) to the theoretical one, the attenuation estimate in
B04 is constrained using many emission lines (although it is dominated
by \ha/\hb).  Using a suite of models, B04 apply a Bayesian approach
to produce the probability distributions for each of the four
parameters (gas metallicity, ionization parameter, dust attenuation,
and dust-to-metal ratio). B04 thus simultaneously produce attenuation
estimate as well as the attenuation-corrected SFR within SDSS $3''$
fiber. SFR within the fiber is constrained by many emission lines,
with the greatest weight carried by \ha. This is one reason why we
will denote B04 SFR and attenuation estimates by ``\ha'', with
quotation marks again serving to indicate that not only \ha\ was
involved in those estimates. The above procedure is directly
applicable {\it only} to galaxies for which the ionizing source is
predominantly star formation, and not, for example, an AGN
activity. B04 therefore calculate {\it fiber} SFRs directly from
emission lines only for galaxies classified in the BPT diagram as star
forming and low-S/N star forming (\S \ref{ssec:bpt}). For other
classes of galaxies (AGN, composite, galaxies with no \ha) for which
either there is no emission line detection or the lines are
contaminated by a non-SF ionizing source, B04 use the relation between
the D4000 spectral index and the specific SFR (SFR normalized by
stellar mass) that has been calibrated using the {\it star-forming}
galaxies. Therefore, for these galaxies, emission-line SFRs are used
indirectly. This is another reason why we denote B04 SFRs as ``\ha''
with quotation marks. We will return to this calibration in \S
\ref{ssec:comp_all}. The above procedure gives only the SFRs within
the $3''$ aperture of SDSS spectroscopic fibers, i.e, the {\it fiber}
SFRs. In order to obtain {\it total} SFRs, an aperture correction is
required. The total SFR is the sum of the SFR within the fiber and
outside of it. To estimate SFR outside of the fiber, galaxies are
first divided into a color-color grid, based on their {\it fiber}
colors. Then, for each color cell, a distribution of $i$
band-luminosity normalized SFRs (SFR/$L_i$) is constructed by adding
up all galaxies of SF or low-S/N SF class that fall within the given
color cell. Then, the color of a galaxy {\it outside} of fiber is used
to select the appropriate color cell, and the SFR/{$L_i$} in that cell
is used with the $i$-band luminosity to derive the SFR outside of
fiber. This is added to previously found SFR within the fiber to
obtain the total SFR. In other words, the aperture correction
procedure is based on two assumptions. First, that SFRs within fibers
have the same dependency on color as the SFR outside of it, and
second, that this dependency, calibrated with star-forming galaxies,
applies to other galaxy classes. The aperture correction factors range
from close to one (no correction) to around a hundred-fold. On average
they are 0.9 dex for the entire sample and 0.6 dex for the
star-forming class.

\subsection{Comparison of ``UV'' and ``\ha'' SFRs for all galaxies} 
\label{ssec:comp_all}

The comparison between SF indicators serves the obvious purpose of
providing a better understanding of each indicator, but also ensures
the mutual check on the reliability of the techniques used to produce
the SFR estimates---in this case, the SFRs based on UV (this study)
and on \ha\ (from B04).

We begin by comparing the formal errors of the two methods. Note that
B04 used a sample based on the expanded SDSS DR1. Here, when we refer
to B04 SFRs we mean the B04 calculations applied to SDSS DR4, and
available as a part of MPA/JHU SDSS value-added catalog. In Table
\ref{tab:errors} we show average formal errors in SFR for the entire
sample, and for each galaxy class. We see that the formal errors of
the two SF estimates happen to be comparable.

\begin{figure}
\epsscale{1.2}
\plotone{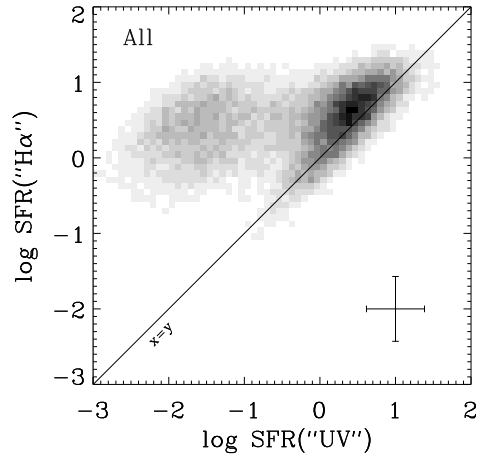}
\caption{Comparison of dust-corrected SFRs derived by \citet{b04}
  (B04, ``\ha'') and the dust-corrected SFRs from this study
  (``UV''). Comparison is given for all galaxies in the sample. Our
  (``UV'') SFRs come from the modeling of the broad-band SED (two
  \galex\ UV bands and five SDSS optical bands), and span almost 5
  orders of magnitude. While most galaxies compare reasonably well,
  those with low ``UV'' SFRs are quite offset from the equality
  line. The error bar represents average formal errors (from Bayesian
  fitting) of the two estimates. Both estimates are given for
  \citep{chabrier} IMF.}
\label{fig:sfr_comp_all}
\end{figure}

\begin{figure}
\epsscale{1.2}
\plotone{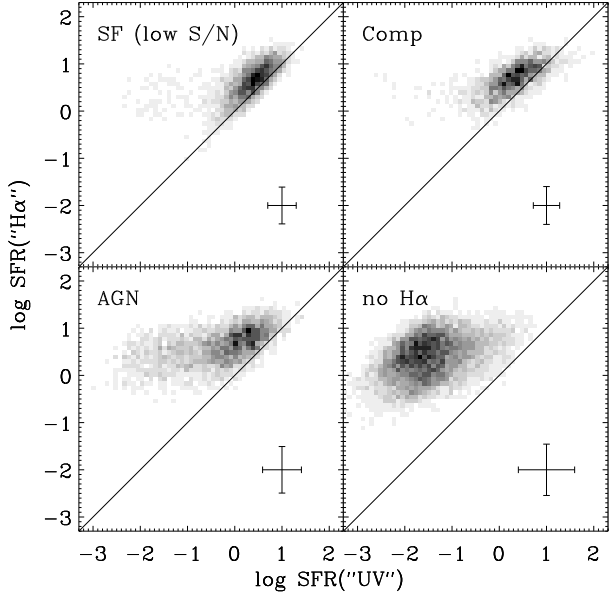}
\caption{Comparison of B04 (``\ha'') and ``UV'' dust-corrected SFRs
  for low-S/N star-forming, SF/AGN composite, AGN, and galaxies
  without \ha\ detection. Note that except for low-S/N SF class, the
  B04 estimate comes from relations calibrated using the ``\ha'' SFRs
  of galaxies classified as star forming. In any class the
  discrepancies are particularly large when the ``UV'' estimate of SFR
  is low. Error bars represents average errors in each class.}
\label{fig:sfr_comp_4classes}
\end{figure}

To facilitate the comparison, we first convert B04 SFRs that were
calculated for \citet{kroupa01} IMF to \citet{chabrier} IMF used in
this study (both with 0.1-100 $M_{\odot}$ limits). From \citet{bc03}
models we find the conversion factor of 1.06 (in the sense that Kroupa
IMF SFRs are slightly higher). In Figure \ref{fig:sfr_comp_all} we
plot B04 ``\ha'' SFR estimates against the ``UV'' for all galaxies in
the sample (throughout the paper SFR always means dust-corrected
SFR). Note that the ``UV'' SFRs exhibit 5 orders of magnitude of a
dynamic range--- from all but negligible rates to those approaching
100 ${\rm M_\odot yr^{-1}}$. We notice that a large number of galaxies
(those with higher ``UV'' SFRs) has comparable SFRs. However, there is
a plume of galaxies for which B04 SFRs are up to 2 orders of magnitude
higher than the ``UV'' ones. This systematic discrepancy cannot be
explained by random errors of the two methods. Recall that B04 do not
derive SFRs directly for classes other than star-forming (and low S/N
star-forming), but instead use relations calibrated on SF class
sample. Thus, in Figure \ref{fig:sfr_comp_4classes} we break the
comparison into galaxy classes (omitting for the moment the SF
class). For the low-S/N star-forming class (upper left panel), for
which B04 SFR estimates do come from the emission lines, the
comparison is generally good for the majority of galaxies (apart from
the overall offset that makes ``\ha'' SFRs higher).  For the other
three classes in Figure \ref{fig:sfr_comp_4classes}, B04 SFRs rely on
calibrations made using the SF galaxies. Composite (SF/AGN) galaxies
compare relatively well, but the comparison becomes significantly
worse for AGN, where we can see both the significant offset for
galaxies with higher SFRs, and the prominent plume for those with
lower ``UV'' SFRs. Finally, for galaxies for which no \ha\ detections
were possible, the discrepancy is very large for almost all
galaxies. It is this class that contributes the most to the plume seen
in the comparison of all galaxies (Figure
\ref{fig:sfr_comp_all}). Given that the galaxies without \ha\
detections belong almost exclusively in the red sequence (Figure
\ref{fig:cmd}), the low SFRs rates estimated by the SED fitting
(``UV'') appear more realistic (see also \S \ref{ssec:ssfr}). Also,
such high SFRs as estimated by B04 would result in a high fraction of
No \ha\ galaxies to have a detectable \ha\ emission, which is
obviously not the case. In general, it appears that the largest
discrepancies between the ``\ha'' and ``UV'' rates occur for galaxies
with low SFRs, especially when B04 estimate SFRs based on calibrations
that employed SF galaxies.

\begin{figure}
\epsscale{1.0} 
\plotone{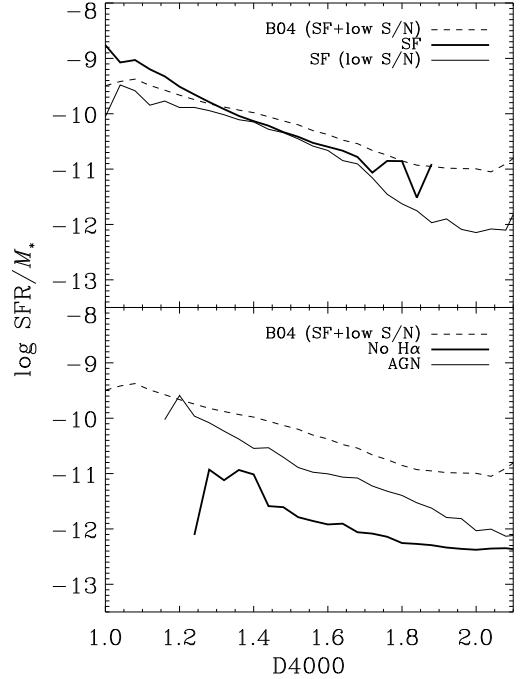}
\caption{Relationships between SFR/$M_*$ (the specific SFR) and the
  D4000 spectral index. In the upper panel we compare the relationship
  from B04 (constructed from SF and low-S/N SF galaxies combined ) to
  our ``UV'' relationships for the SF and the low-S/N SF class,
  separately. The B04 relationship comes from measurements within the
  fiber, while others are total, so this comparison is primary an
  illustration. B04 use such relationship to determine {\it fiber}
  SFRs of non-SF classes of galaxies. Note that beyond D4000 = 1.9,
  the B04 calibration is actually based on low S/N spectra which may
  be contaminated by AGN. In the lower panel the same B04 relationship
  is compared to our ``UV'' rates for the AGN, and for galaxies with
  no \ha\ detection. The two follow different relationships, with AGN
  and ``No \ha'' classes having much lower specific SFRs than the SF
  class. The non-unique mapping between the specific SFR and the D4000
  index for various galaxy groups, and the fact that the correlation
  cannot be established well using the emission lines indicates that
  fiber SFRs in B04 may be systematically affected.}
\label{fig:ssfr_d4000}
\end{figure}

\begin{figure}
\epsscale{1.2} 
\plotone{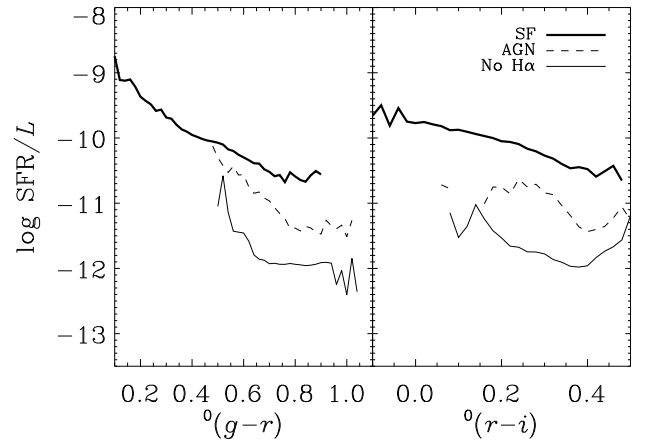}
\caption{Relationship between the luminosity-normalized SFR and
  optical colors. B04 use luminosity-normalized SFRs binned in $g-r$
  vs.\ $r-i$ color-color grid to estimate SFR outside of fiber. They
  build their relationships using SF galaxies, but apply them to other
  types as well. Here, based on ``UV'' measurements, we show that
  different classes of galaxies follow different relationships, and
  that applying the one from SF galaxies would lead to systematic
  differences, which cause most of the SFR discrepancies seen in
  Figure \ref{fig:sfr_comp_4classes}.}
\label{fig:sfrl_color}
\end{figure}

B04 SFRs are the sum of SFRs within and outside of fiber. To trace the
source of the SFR discrepancies, we revisit the calibrations used to
derive B04 SFRs for non-SF classes. First, within the fiber, B04 base
the SFRs for non-SF classes on the relation between the specific SFR
(in the fiber) and the D4000 index calibrated using the SF and low-S/N
SF galaxies. We show this calibration in both panels of Figure
\ref{fig:ssfr_d4000} as the dashed line (this can be compared to their
Figure 11). Then, B04 use D4000 to determine the specific SFR (and
therefore the SFR itself) of galaxies of other classes. In the upper
panel of Figure \ref{fig:ssfr_d4000}, we compare B04 relationship to
our (``UV'') relationships for the star forming (bold solid line) and
the low-S/N star-forming (thin solid line) galaxies. Note that in
principle we cannot compare the B04 specific SFR inside of fiber to
the total specific SFRs obtained from the UV, so the comparison is
more for illustrative purposes. At D4000 $<1.7$ either SF or low-S/N
SF lines from ``UV'' compares well to B04. However, notice that there
are {\it no} galaxies classified as star forming beyond D4000 =
1.9. Therefore, from that point on, B04 relationship is based only on
low-S/N SF class which are most likely just AGN contaminants (see \S
\ref{ssec:bpt}). In contrast to B04, the ``UV'' relationship for low
S/N SF continues to decline as D4000 increases. In the lower panel of
Figure \ref{fig:ssfr_d4000} we show the same B04 calibration against
``UV'' rates for the AGN (thin solid line) and galaxies with no \ha\
detections (bold solid line). In these cases, the relationship for
these two types as indicated by the ``UV'' is entirely different from
that of the star forming galaxies, even at lower D4000 values. For
clarity we omit showing the composite class.  Altogether, this leads
to a conclusion that the relationship between the specific SFR and the
D4000 index is not unique for different classes of galaxies. We tested
if this is an artifact of the fact that D4000 comes from the central
$3''$, while the specific SFR for ``UV'' is integral. We find that the
relationships are qualitatively the same for $z<0.07$ and for $z>0.12$
samples for which D4000 probes different physical sizes. Instead, the
most likely explanation is that since D4000 and specific SFR are
sensitive to SF activity on different timescales, the galaxies with
differing SF histories will have different relationships. To much
lesser extent, some of the differences could be due to metallicity
\citep{poggianti}.

Given the level of aperture corrections (a factor of several), the
analysis given above does not provide the explanation for the large
discrepancies we see when comparing the {\it total} SFRs. Since the
SFR outside of fiber dominates in B04 total SFR, we will now focus on
the calibration B04 use to determine it. As outlined in \S
\ref{ssec:ha} B04 aperture correction relies on the calibration of the
luminosity-normalized SFR as a function of $g-r$ and $r-i$
color. Using our UV-based SFRs we can test one of the two assumptions
behind B04 aperture corrections---whether the SFR/$L$ calibration
against color holds for different classes of galaxies. In Figure
\ref{fig:sfrl_color} we plot SFR/$L$ against two colors that B04 use
to determine SFRs outside of fiber. The bold line represents the
relation from galaxies classified as star-forming. This is the basis
for B04 calibration which they apply to other classes as well. The
dashed and the thin solid lines show the relations for AGN and
galaxies with no \ha. We see (left panel) that as the $g-r$ color
increases, the discrepancy between SF and other classes rises, and
reaches some 0.8 dex for AGN and 1.3 dex for No \ha. Similar levels of
difference are present against $r-i$ (right panel). The level of these
differences matches the discrepancies (plumes) in the comparison of
B04 and ``UV'' SFRs in Figure \ref{fig:sfr_comp_4classes}. We can
interpret these differences as reflecting the fact that the colors
used for B04 calibration trace old populations, while the current SFR
traces young populations, and it is not surprising that the two will
differ for different classes of galaxies. Also note that the
discrepancies of specific SFRs of different classes of galaxies are
qualitatively similar in the case of D4000 and in the case of $g-r$
color, which is not surprising since they are both sensitive to
population age on the similar timescales, and are similarly not
affected by the dust. To conclude, UV-derived SFR is subject to fewer
limitations, so it can be applied to a more diverse types of normal
galaxies.

\subsection{Comparison of ``UV'' and ``\ha'' SFRs for star-forming galaxies} 
\label{ssec:comp_sf}

\begin{figure}
\epsscale{1.2}
\plotone{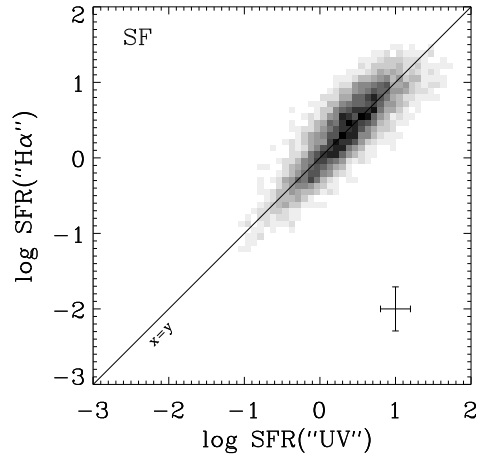}
\caption{Comparison of B04 (``\ha'') to our ``UV'' SFRs (both
  independently dust-corrected) for galaxies classified as
  star-forming. The two compare very well on one-to-one basis. Also,
  the scatter is compatible with each measurement's errors.}
\label{fig:sfr_comp_sf}
\end{figure}

To avoid the problems indicated in the previous section, we need to
compare the SFR estimates from the ``\ha'' and the ``UV'' methods in
galaxies where \ha\ is well-detected (S/N $> 3$) and arises
predominantly from star formation, i.e., to galaxies classified as
star-forming (SF). In Figure \ref{fig:sfr_comp_sf} we show the B04
``\ha'' star formation rates plotted against our ``UV''. We see that
the comparison, spanning some three orders of magnitude in SFR, is
exceptionally good. Formal error bars are comparable, with the ``UV''
being somewhat smaller (see Table \ref{tab:errors}). The scatter
(standard deviation of the difference) of the two measurements is 0.50
dex. When $3\sigma$ outliers are excluded, the scatter is reduced to
0.36 dex. This is very well matched by the sum (in quadrature) of the
formal errors of the two methods (0.35 dex), confirming that the two
measures are predominantly independent. There is an average offset
between the two SFRs of only 0.06 dex in the sense that ``\ha'' SFR is
higher (which reduces to 0.02 dex, i.e., 5\%, when $3\sigma$ outliers
are excluded).

\begin{figure}
\epsscale{1.2}
\plotone{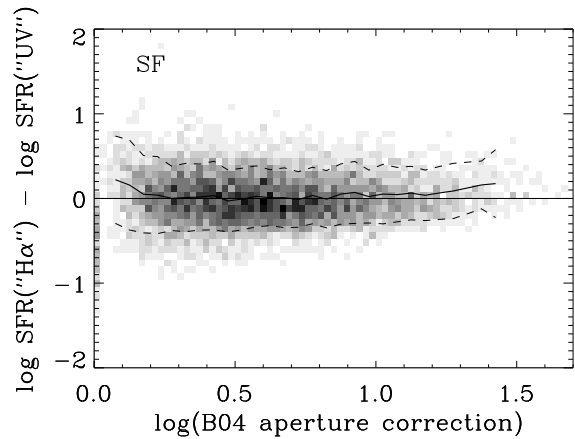}
\caption{The difference of B04 (``\ha'') and our ``UV'' SFRs, for
  galaxies classified as star-forming, as a function of aperture
  correction applied to B04 SFRs. There is no apparent trend,
  suggesting that the aperture corrections derived by B04 are quite
  robust.  Running average (in 0.05 dex bins) is shown as a thick
  line, and the $\pm 1\sigma$ range as dashed lines.}
\label{fig:dsfr_apcor}
\end{figure}

\begin{figure}
\epsscale{1.2}
\plotone{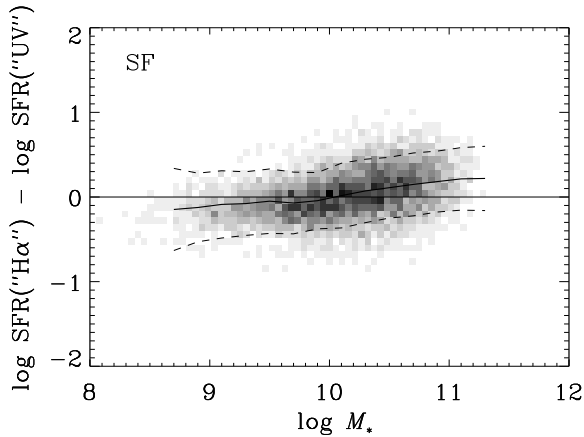}
\caption{The difference of B04 (``\ha'') and our ``UV'' SFRs, for
  galaxies classified as star-forming, as a function of galaxy stellar
  mass. There is a clear trend with respect to mass, leading to a
  difference of SFRs of 0.38 dex (a factor of 2.4) over the $8.5<\log
  M_* <11$ range. Running average (in 0.2 dex bins) is shown as a
  thick line, and the $\pm 1\sigma$ range as dashed lines.}
\label{fig:dsfr_mass}
\end{figure}

However, Figure \ref{fig:sfr_comp_sf} alone can potentially hide some
systematic trends between the two SFR estimates. Where can the
differences in the SFR estimates arise from?  Deriving SFRs for either
methods incorporates several steps: (a) obtaining full ``\ha'' or
``UV'' observed luminosity (which in the case of ``\ha'' involves
aperture corrections), (b) correcting the observed luminosity for dust
attenuation, and (c) converting the luminosity into a SFR. While both
methods perform these three steps simultaneously, we can still design
tests that would differentiate between. Given in some cases the large
aperture correction applied by B04, we first plot in Figure
\ref{fig:dsfr_apcor} the difference in SFR estimates (SFR residuals)
with respect to the level of aperture correction applied to ``\ha''
rates. We see no correlation. We also check, but do not find a
correlation between the residuals and the apparent sizes of galaxies
(not shown). Those are good indications that the aperture corrections
of B04 are quite reliable for galaxies classified as star-forming.
Finally, we find that the SFR residuals exist when plotted against the
stellar mass of a galaxy (shown in Figure \ref{fig:dsfr_mass}). We
have a change of SFR residual of 0.38 dex over the $8.5<\log\, M_* <11$
range, with $1\sigma$ dispersion of 0.33--0.41 dex around the running
mean. To double check if the residuals are in any way connected with
B04 aperture corrections, we limit the sample to farther, and
therefore on average smaller galaxies ($z>0.12$). We still find that
the residuals correlate with stellar mass.

Before ruling out step (a) as being responsible for the mass-dependent
residuals, we perform two additional checks. First, imagine there is
an offset in zero points between \galex\ and SDSS magnitudes (after
all, they come from different surveys and are not measured in
identical apertures). E.g., if \galex\ fluxes were systematically
overestimated, this could cause ``UV'' SFRs to be overestimated as
well, possibly in such a way to preferentially boost blue, low-mass
galaxies (as the trend in Figure \ref{fig:dsfr_mass} would
suggest). To test this, we perform full SED-fitting runs in which we
make $FUV$ and $NUV$ fainter by 0.1, 0.2 and 0.4 mag. We first check
the quality of these new fits by comparing the distribution of
$\chi^2_{\rm best}$ values with the nominal (no magnitude offset)
run. Runs with 0.2 and 0.4 mag offsets produce evidently inferior
fits, already suggesting that any potential offset cannot be that
large. The run with 0.1 mag offset, however, appears as good as the
nominal run, with an even slightly smaller average $\chi^2_{\rm
best}$. However, comparing the ``UV'' SFRs produced with offset
\galex\ magnitudes to ``\ha'' SFRs, we find that while slightly
flattening the slope of the residuals with respect to mass, the trend
is not eliminated. In fact, the scatter of the residuals around the
mean is larger when modified ``UV'' SFRs are used instead of
nominal. We also look at the quality of fits with \galex\ magnitudes
offset in the opposite direction (making them brighter), but such fits
are evidently inferior.

The second test concerns SDSS magnitudes. Namely, B04 use {\sc CMODEL}
SDSS magnitudes to transform fiber SFRs into total SFRs, while we use
{\sc MODEL} SDSS magnitudes when performing SED fitting.\footnote{SDSS
{\sc MODEL} magnitudes (defined as either the exponential or de
Vaucouleurs magnitude, depending on which profile better describes a
galaxy) are preferred magnitudes for SED fitting as they preserve
relative fluxes (colors) better. {\sc CMODEL} magnitudes (defined as a
composite of exponential and de Vaucouleurs magnitude) should provide
a good measure of a total galaxy light.} Therefore, we perform another
SED fitting using {\sc CMODEL} magnitudes instead. First, we notice
that the quality of fitting with {\sc CMODEL} magnitudes is noticeably
inferior, stressing their inadequacy for producing reliable color
estimates. Also, the use of {\sc CMODEL} magnitudes does not remove
the trend of SFR residuals, and the scatter of the residuals becomes
larger than in the nominal run, especially at higher masses. With this
test we exhaust the possibilities that discrepancies arise in step (a)
above.

\begin{figure}
\epsscale{1.2}
\plotone{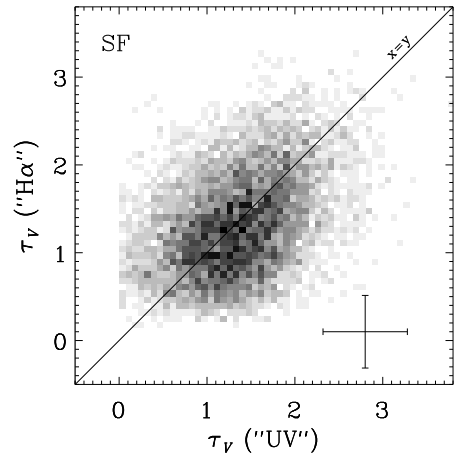}
\caption{Comparison of B04 (``\ha'') to our (``UV'') estimates of
  $V$-band dust opacity (attenuation), for the SF galaxy class. The
  formal error of both estimates (average errors shown) is large,
  leading to a large scatter. However, there is an agreement between
  the two in the general sense. B04 $\tau_V$ is to first order
  constrained by the \ha/\hb\ ratio (Balmer decrement), while our
  estimate is predominately constrained by the UV slope. Both
  estimates were made in accordance with the \citet{cf00}
  two-component dust attenuation model.}
\label{fig:tau_comp}
\end{figure}

\begin{figure}
\epsscale{1.2}
\plotone{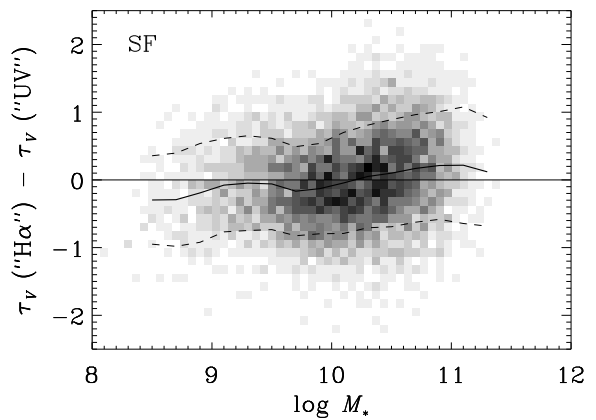}
\caption{The difference of B04 (``\ha'') and our ``UV'' estimates of
  $V$-band dust opacity as a function of galaxy stellar mass (for SF
  galaxies). There is a trend with respect to mass, leading to a $0.5$
  difference over the $8.5<\log\, M_* <11$ range. Running average (in
  0.2 dex bins) is shown as a thick line, and the $\pm 1\sigma$ range
  as dashed lines.}
\label{fig:dtau_mass}
\end{figure}

We now move onto step (b), i.e., correcting the observed flux for dust
attenuation. Note that both B04 and this study use \citet{cf00}
prescription for dust attenuation, the aim of which was to produce
{\it consistent} treatment for \ha, and UV continuum attenuation. Is
the presence of the SFR residuals an indication that this dust
attenuation model does not produce fully consistent answers? In our
SED fitting we keep track of estimates on $\tau_V$---the dust opacity
in rest-frame $V$ band. We can thus compare our $\tau_V$ values with
those obtained by B04 (which we denote $\tau_V$(``UV'') and
$\tau_V$(``\ha''), respectively). While we use the same model to
constrain attenuations, B04 obtain them from the emission lines (to
first order from the Balmer decrement), while we obtain them from the
broad-band SED (to first order from the UV spectral slope, $\beta$, or
equivalently, the UV color). Also, B04 measurements are restricted to
the fraction of the galaxy inside of the fiber, which can in principle
produce some systematic differences. Direct one-to-one comparison,
shown in Figure \ref{fig:tau_comp}, shows a rough agreement, albeit
with a large scatter due to relatively large errors of both estimates
(average formal errors shown as the error bar). Looking instead at the
distributions of the two $\tau_V$ estimates (not shown), we find them
to be very similar on the whole, with the $\tau_V$(``UV'') slightly
offset towards the larger values with respect to $\tau_V$(``\ha'').
Since we find SFR residuals to correlate with the mass, it is more
instructive to check whether such a trend exists in the difference of
$\tau_V$ estimates from ``\ha'' and ``UV''. In Figure
\ref{fig:dtau_mass} we plot $\tau_V$ residuals against the stellar
mass. Despite the large scatter, we find a systematic trend. The
gradient is 0.47 (equivalent to 0.52 mag) over the $8.5<\log\, M_* <11$
range. For lower masses, ``UV'' $\tau_V$ attenuation is higher than
``\ha'', which implies that the observed UV flux is corrected more
than $\tau_V$(``\ha'') would require, resulting in ``UV'' SFRs being
higher than SFR(``\ha''). The situation is reversed at high
masses. This trend matches the sense of the trend of SFR residuals.

\begin{figure}
\epsscale{1.2}
\plotone{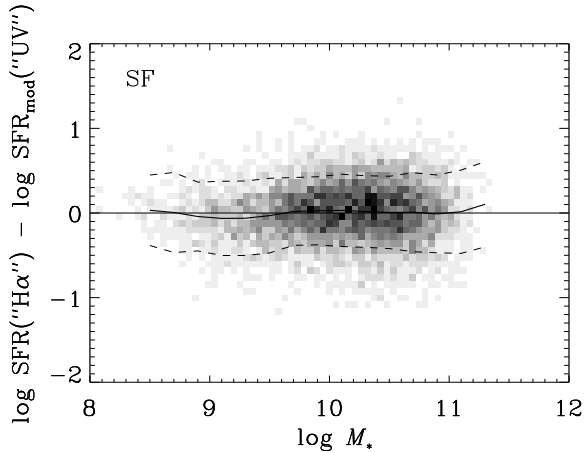}
\caption{The difference of B04 (``\ha'') and the {\it modified} ``UV''
  SFRs as a function of galaxy stellar mass. We modify one of the two
  SFRs (in this case the ``UV'' SFR) by the difference in
  B04 and our attenuation estimate. After this modification, the
  trend of the residuals with respect to mass is eliminated, and the
  overall difference is close to zero. Therefore, we find that
  emission lines and the continuum essentially produce the same SFR
  estimates. Running average (in 0.2 dex bins) is shown as a thick
  line, and the $\pm 1\sigma$ range as dashed lines.}
\label{fig:dsfr_corr_mass}
\end{figure}

In order to establish whether the difference in $\tau_V$ estimates is
the dominant cause of the trend of SFR residuals, we will try to
modify one of the two SFR estimates (e.g., the ``UV'' SFR) based on
the difference in $\tau_V$ values. Note that this is an approximate
technique, since our SED SFRs are not ``corrected'' for dust by the
application of a single number, but rather by a complex application of
\citet{cf00} prescription on many different populations that
constitute the model SED. Nevertheless, for the purposes of this test
we would assume that the attenuation in the FUV drives the correction
of the ``UV'' SFR estimate. We therefore modify it with:

\begin{equation}
{\rm SFR(``UV'')_{mod}} = {\rm SFR(``UV'')} \times 10^{0.4\Delta
  A_{FUV}},
\label{eqn:sfrmod}
\end{equation}

\noindent where $\Delta A_{FUV}$ is the difference between the
attenuation in the rest-frame FUV as implied by $\tau_V$(``\ha'') and
that determined from the SED fitting. We look at the parameters from
the SED fitting to calibrate the relationship between the attenuation
in FUV and the $V$-band opacity, and find the following relationship:

\begin{equation}
\Delta A_{FUV} = \mu(5.4-0.84 \tau_V) \Delta \tau_V,
\label{eqn:deltaafuv}
\end{equation}

\noindent where $\Delta \tau_V= \tau_V\rm{(``H\alpha'')} -
\tau_V\rm{(``UV'')}$, while $\tau_V$ and $\mu$ come from the SED
fitting (i.e., they are ``UV'' measurements). The parameter $\mu$ is
the coefficient in the \citet{cf00} model that determines the fraction
of the total opacity affecting the diffuse ISM. We modify SFR(``UV'')
according to Equations \ref{eqn:sfrmod} and \ref{eqn:deltaafuv}, and
in Figure \ref{fig:dsfr_corr_mass} re-plot the difference in SFRs
against the mass. The trend of the SFR residuals with mass has
disappeared, and the difference stays within $\pm 10\%$ across the
full mass range. The dispersion around the running mean is 20\% larger
than with the non-modified SFR residuals, which is to be expected due
to the approximate nature of the modification applied. Therefore, we
conclude that the ``UV'' continuum and the ``\ha'' in essence produce
identical answers for the star formation rate, the apparent difference
stemming from different estimates of the attenuation that the two
methods provide. With this in mind, we can also rule out other
possibilities for the cause of the original differences in the SFRs,
such as the reliability of converting the luminosity into a SFR (point
(c) above). Also, these results show that, contrary to some notions,
\citet{bc03} models cannot be too much off in the UV.

Although we can now account for the SFR differences, we would still
like to understand the origin of attenuation differences that cause
them. Related to this, we should try to determine which attenuation
estimate (emission-line or the UV continuum) is more accurate. At this
point we need to emphasize that our SED fitting has a non-flat prior
distribution of $\tau_V$ (described in \S \ref{ssec:models}), while
B04 uses a flat prior, with attenuation taking values in the
$0.01<\tau_V<4$ range. While the non-flat prior used in our study is a
reasonable assumption, and the one that helps constrain the parameter
space to physically realistic values, we would like to check whether
the differences in priors induce any systematic effects. To that
effect we first perform an additional SED fitting run, with the model
libraries created with a flat attenuation prior, taking values in the
$0<\tau_V<3$ range. At lower ($\tau_V<0.7$) values, we find the
distribution of the resulting $\tau_V$ values to be similar to the
original distribution, while at higher values it displaces the
original peak, located at $\tau_V\approx 1.3$, to a broader
distribution peaking at $\tau_V\approx 2.0$. The latter is the obvious
consequence of assigning equal probability to models with high
attenuations. In any case, we end up with a $\tau_V$ distribution that
differs more with respect to the B04 $\tau_V$(``\ha'') than did the
original $\tau_V$(``UV''). Nevertheless, we proceed and check the
relationship between the new SFR residuals and the stellar mass.  The
trend observed in the original relationship is still present, although
it is somewhat weaker at low masses. Anyhow, it appears that we cannot
force the two attenuation estimates to reach an exact agreement by a
simple modification of the prior distribution used in SED fitting.

\begin{figure}
\epsscale{1.0}
\plotone{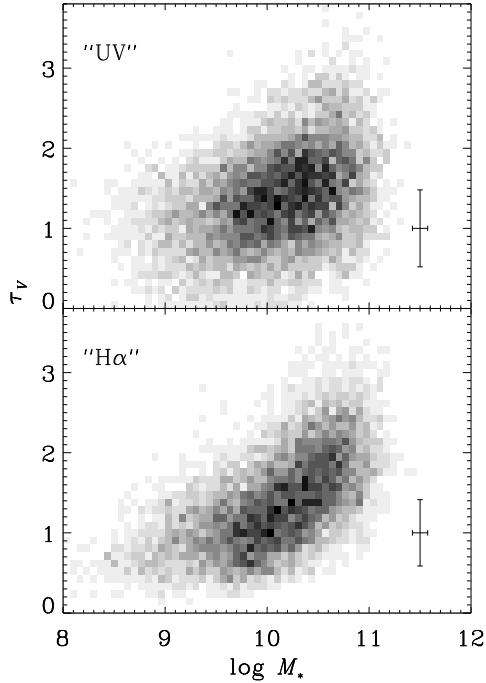}
\caption{Attenuation-mass relationship for star-forming galaxies. The
  two attenuation estimates, one from our SED fitting (``UV'', upper
  panel), and the other from B04 (``\ha'', lower panel) are plotted
  against the stellar mass. The B04 estimate (principally from
  \ha/\hb\ ratio) produces a tighter relationship with a greater span
  of attenuation values, indicating that it is probably more
  accurate.}
\label{fig:tau_mass}
\end{figure}

Next we will try to get a sense of which $\tau_V$ estimate is more
realistic. Since the direct comparison with an external independent
measurement is not readily available, we will try to evaluate the
differences using internal relations. For star-forming galaxies we
expect attenuation to be correlated with the stellar mass.  In Figure
\ref{fig:tau_mass} we show this relation for our ``UV'' estimates
(upper panel) and B04 ``\ha'' estimates (lower panel). It is apparent
that the latter defines a more tight relationship. Also,
$\tau_V$(``\ha'') spans a larger range of values than $\tau_V$(``UV'')
(1.4 vs.\ 0.8 over $8.5<\log\, M_* <11$).  Finally,, the dispersion
around a running mean is some 20\% smaller for $\tau_V$(``\ha''). This
all indicates that B04 $\tau_V$ estimates are probably more
accurate. While the two $\tau_V$ estimates agree well on the whole
(Figure \ref{fig:tau_comp}), the ``UV'' estimate appears more poorly
constrained, which may cause it to more often take values away from
the extrema, suppressing the span in relationship with the mass. As a
result, the attenuation estimates for low-mass galaxies will be
overestimated, and for high-mass galaxies underestimated. It is
difficult to directly test this explanation. Namely, if we restrict
the analysis only to galaxies with small formal errors in
$\tau_V$(``UV'') (or alternatively small errors in $FUV-NUV$), we at
the same time bias the sample to smaller values of
$\tau_V$(``UV''). Alternative explanation is that the \citet{cf00}
$\propto \lambda^{-0.7}$ extinction curve deviates from the true
extinction curve in a way that would be mass-dependent (or some
quantity related to mass, such as the metallicity). A possibility that
the extinction curve is mass-dependent is raised in \citet{ben}. We
would be able to acquire further insight by comparing both attenuation
estimates to some external measure, such as the IR excess
(IRX). \citet{ben} have recently obtained IRX attenuations from {\it
Spitzer} MIPS and \galex\ observations of SDSS galaxies, therefore
such sample can be used to include UV (i.e., SED) derived
attenuations. Also, the SED-based attenuation estimates can possibly
be improved by adding the near-IR data, such as the $JHK_s$ photometry
from 2MASS survey \citep{2mass}, which would improve attenuation
estimates by placing stronger constraints on the stellar metallicity.

\section{Obtaining the UV SFRs without SED modeling} \label{sec:recipe}

In this work we are using sophisticated SED modeling to obtain
UV-based star formation rates. In many applications such a detailed
approach is not practical, or not even possible. In such cases one
would like to obtain a reasonably good estimate of a dust-corrected
star formation by applying some simple transformations to the UV
photometry. While such methods have been used in many previous
studies, here we will use the results of the detailed SED analysis to
{\it calibrate} such simplified models, with a special emphasis on
users of UV data obtained with \galex. Since for the foreseeable
future \galex\ will remain the only facility capable of observing a
large number of galaxies in the UV, it is not without justification to
treat its filters as defining some standard photometric bands in the
UV domain. Researchers who study high-$z$ galaxies in rest-frame UV
can calibrate their blueshifted filters against \galex\ FUV and NUV
response curves.\footnote{\galex\ filter response curves can be
obtained from \\
\url{http://galexgi.gsfc.nasa.gov/tools/Resolution\_Response}.}

Obtaining a UV-based SFR consists of K-correcting the UV magnitudes,
estimating the dust attenuation of the FUV flux, and converting the
dust-corrected FUV luminosity into a SFR. We assume that the user will
correct their data for the Galactic extinction, and then apply some
standard K-correction procedure to obtain rest-frame $FUV$ and $NUV$
magnitudes. Of course, obtaining reliable K corrections requires
optical or near-IR photometry. To calibrate the relations in this
section, we use {\sc KCORRECT V4\_1\_4} (\S \ref{ssec:kcor}).

\begin{figure}
\epsscale{1.2}
\plotone{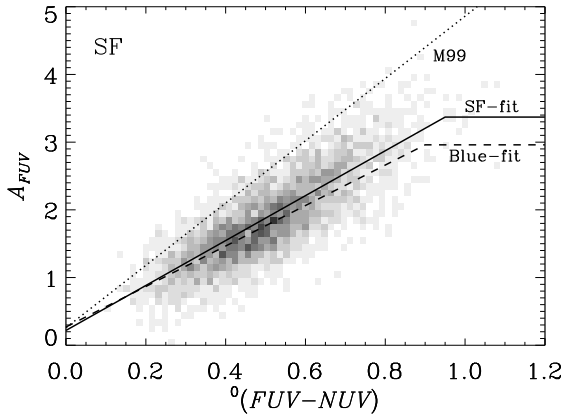}
\caption{Relationship between the attenuation in the \galex\ far-UV
  and the UV slope. The FUV attenuation estimate comes from our SED
  fitting, and the $FUV-NUV$ rest-frame color serves as an indicator
  of the UV SED slope. Data and the solid line linear fit is for
  galaxies classified as star forming. The dashed line is a fit to
  blue (rest frame $NUV-r<4$) galaxies. The fits reproduce the trends
  in general, however, there is a significant scatter.
  \citet{meurer99} relation for starburst galaxies is plotted as the
  dotted line (M99). It is obviously not appropriate for our sample of
  mostly normal star-forming galaxies.}
\label{fig:fuv_atten}
\end{figure}

We start by using the results of the full SED fitting to calibrate the
well-known correlation between the attenuation and the UV spectral
slope (e.g., \citealt{calzetti94}). This is often refereed to as
IRX-$\beta$ relation, although strictly speaking IRX indicates IR
excess, which is {\it correlated} with UV attenuation. IRX-$\beta$
relationship for {\it normal} galaxies, such as those in our sample,
has been previously studied with \galex\ data by
\citet{seibert,cortese,gildepaz,boissier,panuzzo}. In Figure
\ref{fig:fuv_atten} we show FUV attenuations ($A_{FUV}$) obtained from
our SED modeling, for galaxies classified as star-forming, plotted
against the rest-frame UV color, which is linearly correlated to the
UV spectral slope. We see that the majority of objects lies along the
ridge. This confirms that there exists an ``IRX-$\beta$'' relationship
for normal galaxies, not just for starburst galaxies as usually
assumed. To quantify the relationship for our sample, and thus allow
$A_{FUV}$ to be determined from \galex\ observations of normal
galaxies, we fit a linear function to running medians (with value from
each 0.05 mag bin weighted equally). Using the medians is necessary in
order to avoid the fit to be affected by numerous outliers. Also, we
find that past some red UV color the attenuation does not seem to
rise, so we adopt a constant value. In any case, $A_{FUV}$ exhibits a
large scatter for such red colors.

\begin{equation}
A_{FUV} = \left\{\begin{array}{ll}
           3.32\, ^0(FUV-NUV)+0.22, &  ^0(FUV-NUV)<0.95 \\
	   3.37, &  ^0(FUV-NUV)\geq0.95, \end{array} \right.
\label{eqn:afuv1}
\end{equation}

\noindent where $A_{FUV}$ is in magnitudes, and superscript 0
designates rest-frame colors. This relation is plotted in Figure
\ref{fig:fuv_atten} as a solid line. In many applications, especially
at higher redshift, spectroscopic classification may not be
available. In those cases one can select star-forming galaxies based
on their color. For blue-sequence galaxies ($^0(NUV-r)<4$, and without
applying any class selection) one should use a slightly modified
relation:

\begin{equation}
A_{FUV} = \left\{\begin{array}{ll}
           2.99\, ^0(FUV-NUV)+0.27, &  ^0(FUV-NUV)<0.90 \\
	   2.96, &  ^0(FUV-NUV)\geq0.90, \end{array} \right.
\label{eqn:afuv2}
\end{equation}

\noindent shown in Figure \ref{fig:fuv_atten} as a dashed line. Note
that these relationships are optimized for \galex\ bandpasses {\it
and} for normal star-forming galaxies. We overplot \citet{meurer99}
relation as a dotted line (we used \citealt{seibert} transformation to
obtain $FUV-NUV$ from the UV spectral slope). Apparently, this
relation, constructed from a sample of starburst galaxies, is well
above the majority of galaxies in our sample. \citet{seibert} and
\citet{cortese}, using smaller samples (RC3 and cluster galaxies,
respectively) also find that \citet{meurer99} relation overpredicts
FUV attenuation. \citet{seibert} derive a relation that exactly
bisects our SF fit and the \citet{meurer99} relation. Also note that
these relations pertain to the optically selected sample (since
practically all SF galaxies are detected in UV bands, there is no
additional UV selection). In general, samples selected at different
wavelengths will have different IRX-$\beta$ relations, as demonstrated
by \citet{buat05}, using \galex\ NUV and {\it IRAS} 60$\mu$m selected
samples.

For a source with known redshift, the dust-corrected, rest-frame $FUV$
is converted into a FUV luminosity. The final step requires converting
the luminosity into SFR. Following B04 notation, we define:

\begin{equation}
\eta^{0}_{FUV} = L^{0}_{FUV}/{\rm SFR(``UV'')},
\end{equation}

\noindent to be the inverse conversion factor between a dust-corrected
rest-frame FUV luminosity (in erg s$^{-1}$ Hz$^{-1}$) and the
SED-derived SFR. The conversion factors comes from the stellar
population modeling that was used to perform the SED fitting. Like the
equivalent conversion factor for \ha\ luminosity, $\eta^{0}_{FUV}$ is
sensitive to metallicity, albeit more weakly so. For our sample (the
metallicity of which is on average $0.8 Z_{\odot}$), the median
conversion factor is:

\begin{equation}
\log \eta^{0}_{FUV} = 28.165.\label{eqn:conv1}
\end{equation}

The above factor is given for \citet{chabrier} IMF, which is used in
this paper. The often-used conversion factor given by
\citet{kennicutt} for the \citet{salpeter} IMF is:

\begin{equation}
{\rm SFR}(M_{\odot} {\rm yr^{-1}}) = 1.4 \times 10^{-28} L_\nu ({\rm
  erg^{-1} s^{-1} Hz^{-1}}).\label{eqn:rob}
\end{equation}

\noindent We find using \citet{bc03} models that the appropriate
transformation factor between UV-derived SFRs that assume Chabrier and
Salpeter IMFs is 1.58 (both with 0.1-100 $M_{\odot}$
limits)\footnote{Note that the conversion between the IMFs is not
generally the same for \ha-derived SFR, UV SFR, or stellar
mass.}. Thus our ``empirical'' conversion factor (Equation
\ref{eqn:conv1}) for Salpeter IMF becomes:

\begin{equation}
{\rm SFR}(M_{\odot} {\rm yr^{-1}}) = 1.08 \times 10^{-28} L^{0}_{FUV}({\rm
  erg^{-1} s^{-1} Hz^{-1}}). \label{eqn:conv2}
\end{equation}

\noindent This implies that the \citet{kennicutt} conversion factor
(Equation \ref{eqn:rob}) is 30\% higher. We verify that the effect of
the difference in bandpasses (\citealt{kennicutt} factor was given for
the 1500--2800 \AA\ range, while the \galex\ FUV filter spans
1300--1800 \AA) is completely negligible. The actual reasons for the
difference are two-fold. First, even when obtaining the {\it model}
conversion factor using \citet{bc03} and the {\it same} assumptions as
given by \citet{kennicutt} (i.e., solar metallicity and a constant
star formation history), we still find \citet{kennicutt} conversion to
be 15\% higher (i.e., the corresponding $\eta$ is lower than one
produced by \citealt{bc03}). Further, we have that our sample has
average metallicity somewhat lower than the solar. This accounts for
another 5\% difference. Finally, the remaining 10\% difference stems
from the fact that our sample has a variety of star formation
histories, giving $\eta$ that is on average higher than that for
constant SF history. To summarize, for optically selected samples
similar to ours, we suggest using ``empirical'' conversion given in
Equation \ref{eqn:conv1} or \ref{eqn:conv2}.

\begin{figure}
\epsscale{1.2}
\plotone{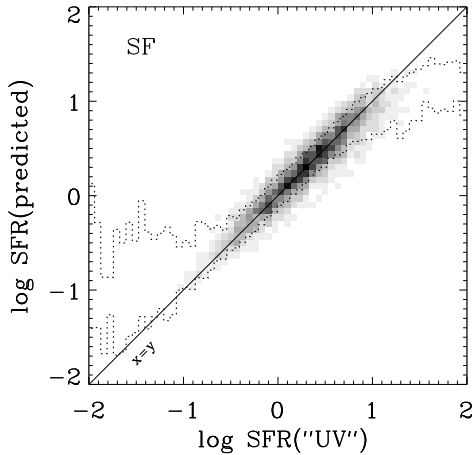}
\caption{Comparison of the SFRs ``predicted'' using a set of simple
  transformations, to SFRs derived using the full SED fitting. The
  predicted SFR is obtained using the transformations that we
  calibrate from the SED SFRs. They are applied to K-corrected $FUV$
  and $NUV$ magnitudes (see \S \ref{sec:recipe}). The comparison is
  good for the majority of galaxies (greyscale scatter plot). The
  dashed lines show 16 and 84 percentile values of the conditional
  distribution at each SFR(``UV''), from which we see that the
  predicted SFRs deviate from SFR(``UV'') at low and high SFRs.}
\label{fig:sfr_pred}
\end{figure}

Finally, we combine the above ingredients and ``predict'' the SFR from
K-corrected $FUV$ and $NUV$ magnitudes. In Figure \ref{fig:sfr_pred}
we compare the SFR obtained using this simple prescription to the
original SFR(``UV'').  We observe a good agreement for the majority of
objects (the greyscale scatter plot). In addition, for each bin in
SFR(``UV'') we show a 16--84 percentile range of the predicted SFR
values, i.e., their conditional distribution. This allows us to track
the trend in the regions where there are few points to show on the
greyscale scatter plot. We see that the predicted SFR starts to
deviate at both the low and the high SFR ends, by up to 1 dex. Good
agreement is obtained in the $-1<\log\, {\rm SFR} < 1$ range, i.e.,
where the majority of galaxies lie. The reason for the deviations at
low and high-SFR end stem from the inability to estimate $A_{FUV}$
reliably only from $FUV$ and $NUV$ for galaxies having extreme SF. No
simple modifications of Equations \ref{eqn:afuv1} and \ref{eqn:afuv2}
are able to reproduce $A_{FUV}$ estimate of galaxies with extreme SFRs
without destroying the agreement for the majority of normal
galaxies. In $-1<\log\,{\rm SFR} < 1$ range, the 16--84 percentile
range is 0.35 dex wide, indicating that the predicted values have an
equivalent $1\sigma$ scatter of 0.17 dex. Therefore, the total error
of the predicted SFRs in this range is 0.26 dex (with the error of
SFR(``UV'') itself added in quadrature). This is a small penalty
considering the ease of the method.  Note that the prescription given
here is optimized to produce good results for the large majority of
galaxies, and that a result for any given galaxy should be used with
caution.

\section{Star formation in the local universe} \label{sec:local}

\subsection{Star formation history and stellar mass} \label{ssec:ssfr}

It is now commonly accepted that the star formation history of a given
galaxy depends strongly on its mass (for early work based on UV see
\citealt{boselli,gavazzi}), and to some extent its environment. To
first order, we can characterize the star formation history using the
specific star formation rate, which is defined as the ratio of the
current SFR to the current stellar mass. Thus, higher values of the
specific SFR indicate that a larger fraction of stars was formed
recently. The unit of specific SFR is inverse time, and it is thus
often referred to as the galaxy build-up time. However, this should
not be confused with an actual galaxy age, since the specific SFR
tells us only how long it would have taken to build a galaxy assuming
it had a current SFR throughout its lifetime, and assuming that the
current stellar mass is close to the total mass formed in stars during
a galaxy's lifetime (the two are not the same because of gas
recycling).

\begin{figure}
\epsscale{1.2}
\plotone{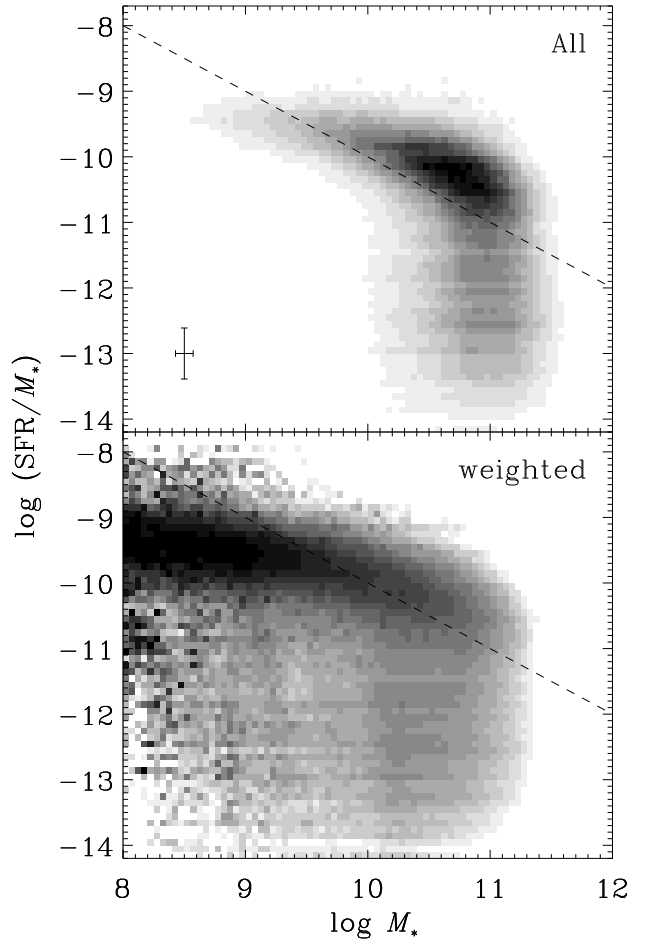}
\caption{Dependence of the star formation history on the stellar
  mass. We use the specific SFR (SFR/$M_*$) as an indicator of a star
  formation history. Galaxies with a larger fraction of recent star
  formation will have a higher value of the specific SFR. Instead of a
  single value, each galaxy is represented with a full two dimensional
  probability function. The dashed line, shown for reference,
  represents a constant SFR of $1\, M_\odot {\rm yr^{-1}}$. The upper
  panel gives equal weight to every galaxy in the sample, while the
  lower panel shows logarithms of volume-corrected values
  (i.e.,weighted by $V_{\rm max}$.) Uneven behavior at low masses is
  because of a small number of galaxies (or no galaxies) in some
  bins. Note that the sample is only optically selected. Our
  completeness limit is below the lowest mass on the plot.}
\label{fig:ssfr_mass_all}
\end{figure}

\begin{figure}
\epsscale{1.2}
\plotone{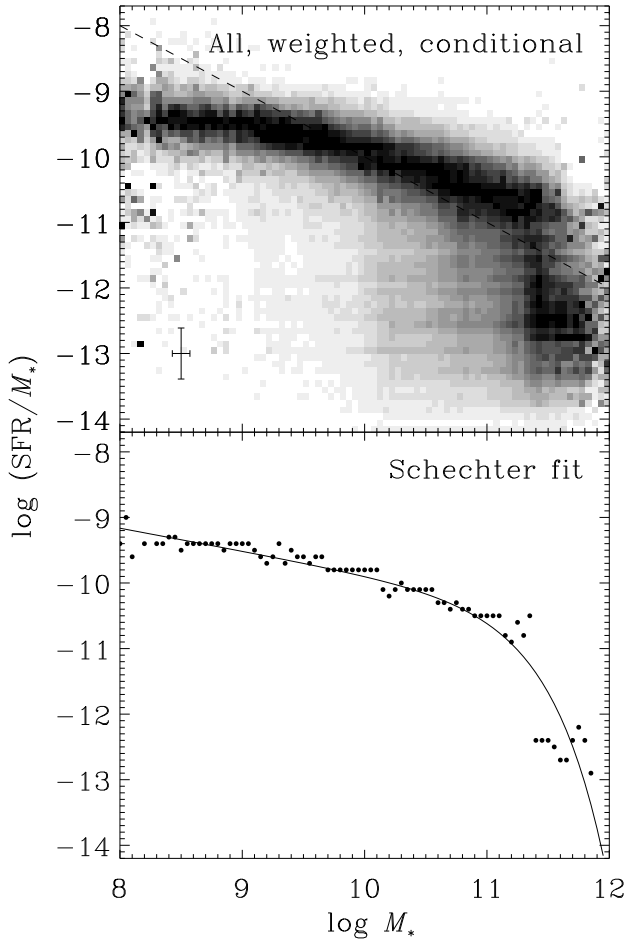}
\caption{Conditional dependence of the star formation history on the
  stellar mass. Same sample as in Figure \ref{fig:ssfr_mass_all}, but
  with each 0.05 dex wide mass bin normalized to its maximum
  separately. This allows us to see what specific SFRs dominate at
  each mass, and to follow activity where there are to few objects to
  show in the standard plot. The lower panel shows running modes from
  the upper panel together with a Schechter-like fit to those points}
\label{fig:ssfr_mass_cond_all}
\end{figure}

For each galaxy in our sample we construct a two-dimensional
probability distribution function in $(\log\,({\rm SFR}/M_*),\log\,
M_*)$ plane as outlined in \S \ref{sec:sed}. We use full 2-D PDF
instead of collapsing it into a single value in order to better
represent low-SFR galaxies whose PDFs are often very wide (i.e., not
well constrained) in specific SFR, and in which cases the single
value, such as the average of the marginalized PDF, would tend to
artificially concentrate in the middle of the PDF. In Figure
\ref{fig:ssfr_mass_all} we show the co-added 2-D PDFs in 0.1/0.05 dex
wide bins. In the upper panel of Figure \ref{fig:ssfr_mass_all}, as in
all previous figures in the paper, we show unweighted values, i.e.,
each galaxy in our sample contributes equally to the greyscale
density. Given our redshift and magnitude limits, we adopt the
$r$-band absolute magnitude limit of $M_r=-14.75$ (which we verify by
constructing a luminosity function). This luminosity limit leads to
the completeness in stellar mass of $\log\, M_* \approx 7.5$, which is
below the plotted mass range. The dashed line is shown for reference
and represents a constant SFR of $1\, M_\odot \rm {yr^{-1}}$. We see
that lower-mass galaxies appear to be confined in a relatively narrow
sequence which declines as the mass increases. At masses above
$10^{10}\, M_{\odot}$, some galaxies assume a much lower value of the
specific SFR, i.e., the sequence turns over. In the lower panel, each
galaxy's PDF is weighted by its $V_{\rm max}$ value, i.e., the sample
is volume-corrected. Since the volume-corrected values span a very
wide range of values, we now show logarithms of number densities,
scaled so that the full greyscale range displays densities from
maximum (black) to 2 dex below the maximum value (lightest shade of
gray). Each horizontal slice in effect represents a mass function of
galaxies of a given specific SFR. It now becomes apparent that the
sequence seen in the unweighted plot extends to lower masses, with an
increasing number density of galaxies. Also, the mass distributions at
low specific SFR extends to lower masses than implied by the
non-corrected plot. Considering the size of the sample, the accuracy
of SFRs, and the sensitivity to low levels of SF, this figure
represents the best available assessment of the star formation history
in the local universe. The typical error in stellar mass is only 0.07
dex, and 0.40 dex in the specific SFR (note that the errors in
specific SFR vary by a factor of three across the sample). We
complement the above figure with Figure \ref{fig:ssfr_mass_cond_all}
(upper panel), where we display {\it conditional} probability
distribution in each 0.05 dex wide mass bin, which is equivalent to
normalizing each column to its maximum value. Such representation
allows us to see what specific SFR dominates at a given mass, and it
also displays more clearly the behavior at masses where there are too
few objects to show on the standard plot. At masses below $10^{10}\,
M_{\odot}$, high specific SFRs dominate. At $M_*=10^{10}\, M_{\odot}$,
more galaxies assume values below the dominant high-specific SFR
``sequence'', down to very low levels. At $M_*=2\times10^{11}\,
M_{\odot}$, close to where the high-specific SFR ``sequence''
terminates, the galaxies with {\it low} specific SFRs begin to
dominate. Above $\log\, M_*=11.8$ the plot is dominated by noise from
very few objects with such high mass. This figure compares well to
Figure 24 in B04.

\subsection{The star-forming sequence} \label{ssec:ssfr_sf}

\begin{figure*}
\epsscale{1.2} 
\plotone{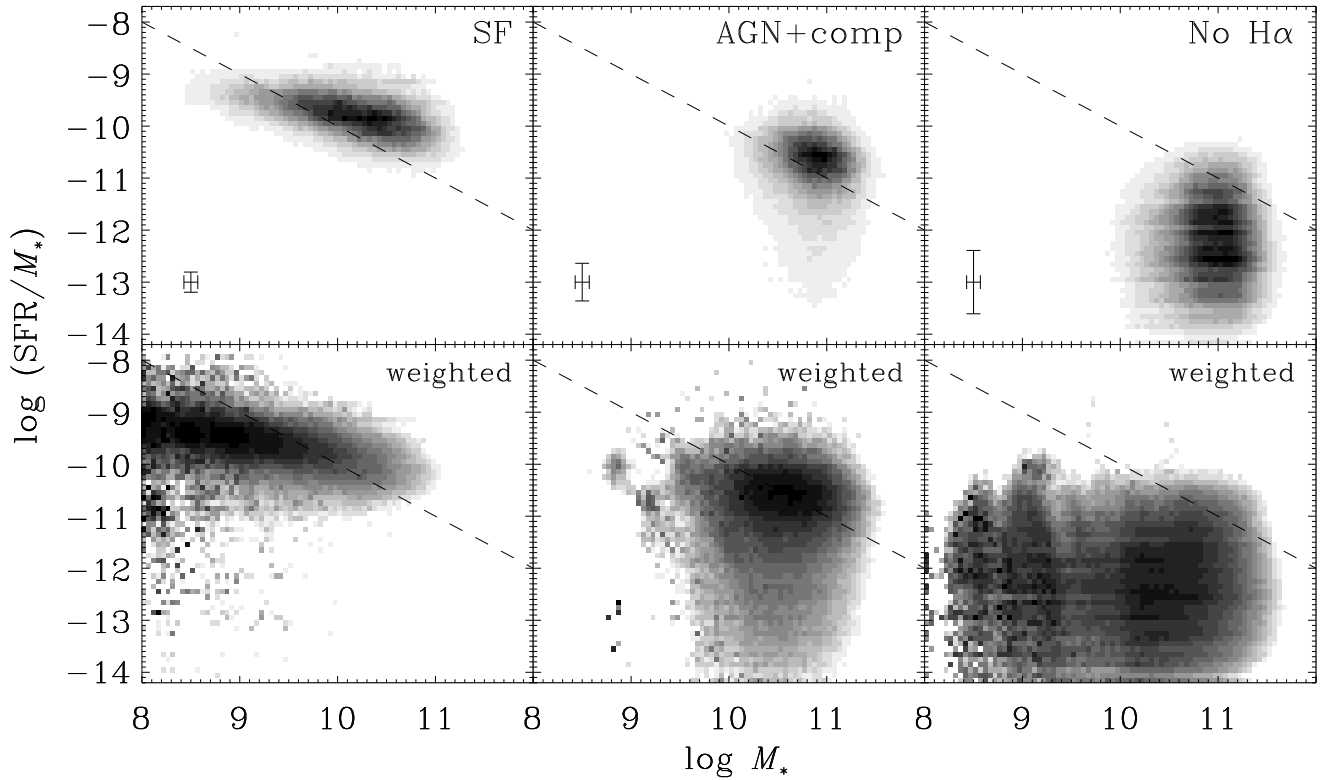}
\caption{Dependence of the specific SFR on the stellar mass for
  different classes of galaxies. Star forming (left panels); galaxies
  with AGN (middle panels); and galaxies without \ha\ detection (right
  panels) all occupy distinct regions of the parameter space,
  indicating different SF histories. SF galaxies form a narrow
  sequence. AGN have intermediate specific SFRs, and are predominantly
  high mass. Galaxies without \ha, mostly red-sequence galaxies, have
  low specific SFRs. The dashed lines shows the reference SFR of $1\,
  M_{\odot} {\rm yr^{-1}}$. The lower panel shows values weighted by
  $V_{\rm max}$. Uneven behavior at low masses is because of small
  number of galaxies (or no galaxies) in those mass bins.}
\label{fig:ssfr_mass_class}
\end{figure*}

We now focus on star formation histories of various classes of
galaxies. In Figure \ref{fig:ssfr_mass_class} we show specific SFR
against the stellar mass for the star-forming, AGN (together with
SF/AGN composites), and the class without \ha\ detection. For each
class the upper panels show nominal, unweighted data, while the lower
panels are volume-corrected.  Dashed line has the same meaning as in Figure
\ref{fig:ssfr_mass_all}. First we notice that the three classes occupy
relatively distinct portions of the parameter space. This is
especially pronounced in the unweighted plots. Thus the three classes
appear to have had quite different star formation histories. This was
to some extent indicated in the CMDs (Figure \ref{fig:cmd}), but is
more striking now.

The star forming class (SF), which forms a blue sequence in the CMD
(Figure \ref{fig:cmd}), dominates the high values of the specific
SFR. The sequence that was obvious in Figure \ref{fig:ssfr_mass_all}
stands completely isolated here. In the weighted plot we see an
ever-increasing density towards the lower mass, which reflects a
rising low-mass slope of the mass function of late type (blue)
galaxies. While the typical SFR {\it rises} from $0.1\, M_\odot {\rm
yr^{-1}}$ at the low-mass end to $10\, M_\odot {\rm yr^{-1}}$ at
$10^{11}\, M_{\odot}$ (consistent with the assumption that the more
massive galaxies contain more gas and therefore have higher SFRs), the
specific SFR actually declines by a factor of 10. The tightness of the
``star forming'' sequence (note that the formal error of the specific
SFR is only 0.20 dex) represents an important indication that the
galaxy's mass regulates the overall star formation history (see also
\citealt{gavazzi2,boselli}). It appears reasonable to assume that in
the absence of an event that may disturb galaxy's gas reservoir, a
galaxy would ``naturally'' sit on this tight sequence. This appears
contrary to the suggestions that the red (non-SF) sequence forms by
simple gas exhaustion, since we would then have a much wider range of
(specific) SFRs at the given mass, for galaxies observed in the
various stages of gas exhaustion. Note that for masses below $\log\,
M_* = 9.5$ we have an additional tail of high specific SFRs (best seen
in volume-weighted plot), which is indicative of {\it starbursts}. We
will return to the width of the SF sequence in \S \ref{ssec:quan}.

\subsection{Specific SFRs and stellar masses of galaxies with an AGN}
\label{ssec:ssfr_agn}

We now turn to galaxies that lie on the AGN branch of the BPT
diagram. These represent both galaxies classified as SF/AGN
composites, and ``pure'' AGN (those lying in the region of the BPT
diagram where non-SF ionizing source is necessary). We have already
seen that a significant fraction of AGN hosts populates the green
valley of the UV-to-optical CMD (Figure \ref{fig:cmd})---the
intermediate region between the blue and the red sequences. (However,
note that in the optical CMD they would lie predominantly in the red
sequence). A possible reason for the presence of these galaxies in the
green valley is that they are entering the stage in which their star
formation is being shut off, and the colors are becoming red as a
result. \citet{chris2}, using the \galex\ data, has demonstrated that
if the SF of galaxies lying in the green valley is indeed being
quenched, then their color evolution implies a mass flux (transfer
from blue to red) that roughly matches the mass increase required to
populate the red sequence between the epochs $z\sim 1$ and $z \sim 0$
\citep{bell,sandy}.

In the middle panels of Figure \ref{fig:ssfr_mass_class} we present
specific SFR against mass for AGN and the SF/AGN composites combined.
The typical error of the specific SFR is 0.35 dex. Galaxies with AGN
populate the region {\it between} the SF galaxies with no AGN and the
galaxies with no \ha\ (adjacent panels), in a more striking manner
than even in the UV-to-optical color-magnitude diagram. We note that
these AGN are of Type 2 (Seyfert 2s and LINERs), which are not
expected to affect the stellar continuum, especially not in the
integrated galaxy light (\citealt{kauff_agn}, and reference
therein). In galaxies having a luminous Type 2 AGN, \citet{kauff_agn}
limit the AGN continuum contribution to less than a few percent in $r$
band. Assuming a power-law AGN continuum of the form $\propto
\nu^{-1.5}$, this limit corresponds to less than $\sim 15\%$
contribution in the UV. Another indication that the UV contribution
from an AGN is low in the UV comes from the fact that the quality of
our SED fitting for galaxies with AGN is nearly the same as for the
similar galaxies with no AGN. Also, \citet{kauff_galex} explore
\galex\ imaging of a more nearby sample of AGN hosts and find that
their UV emission is quite extended, thus unlikely to emanate from the
central AGN.

\begin{figure}
\epsscale{1.2} 
\plotone{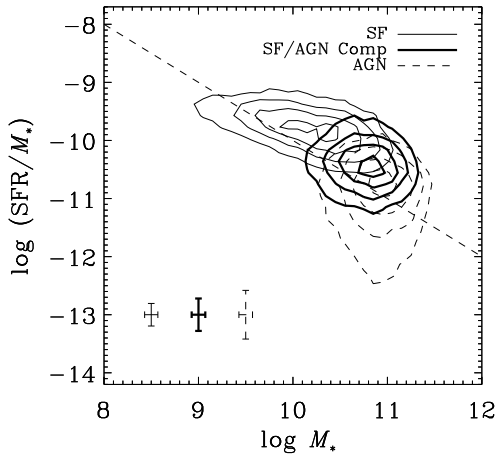}
\caption{Specific SFR and mass of star forming, and of galaxies
  hosting AGN. We plot density contours for normal SF galaxies having
  no AGN (thin solid contours), of SF/AGN composites (bold solid
  contours), and of ``pure'' AGN (dashed contours). The SF/AGN
  composite class (lower part of the AGN branch in BPT diagram)
  bridges normal SF galaxies and the AGN lying on the upper part of
  the AGN branch. Contours of unweighted distribution encompass 10,
  30, 50, and 70\% of objects, i.e., their composite PDF densities.}
\label{fig:ssfr_mass_agn1}
\end{figure}

\begin{figure}
\epsscale{1.2} 
\plotone{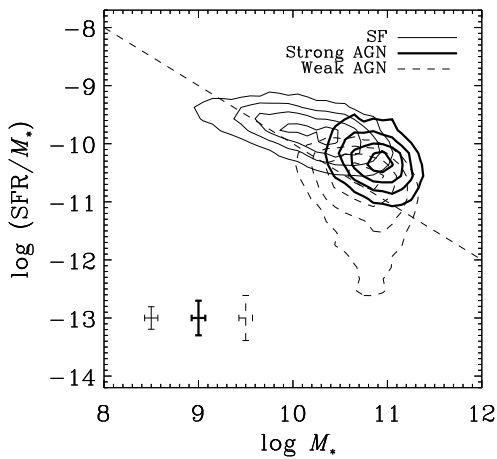}
\caption{Specific SFR and mass of star forming, and of galaxies
  hosting AGN of various strengths. We again show SF galaxies with no
  AGN (thin solid contours), galaxies with strong AGN (bold solid
  contours), and galaxies with weak AGN (dashed contours).  The
  strong/weak division of all galaxies on the AGN branch of the BPT
  diagram is based on the [OIII]$\lambda$5007\AA\ luminosity. Strong
  AGN appear to form a continuation of the SF sequence, having similar
  SFRs, but larger masses. Weak AGN have similar masses as strong AGN
  but lower (specific) SFRs, with a tail extending to values typical
  of "quiescent" galaxies. Contours of unweighted distribution
  encompass 10, 30, 50, and 70\% of objects, i.e. their composite PDF
  densities.}
\label{fig:ssfr_mass_agn2}
\end{figure}

The fact that most AGN have intermediate star formation histories, and
that this region is {\it not occupied} by galaxies unless they have an
AGN, is remarkable in the light of recent suggestions that the AGN
feedback may be regulating the star formation of massive galaxies
\citep{springel}. Still, we would like to know if there is a
connection of galaxies hosting AGN with normal star-forming galaxies
with no AGN. In Figure \ref{fig:ssfr_mass_agn1}, we plot SF class
(thin solid contours), and separate AGN into ``pure'' AGN (dashed
contours) and SF/AGN composites (bold solid contours). Composites form
the lower part of the AGN branch in the BPT diagram, and the ``pure''
AGN the upper part. Contours encompass 10, 30, 50 and 70\% of each
class's objects (i.e., their PDF densities), with no $V_{\rm max}$
weighting. We notice that composites ``bridge'' the normal SF
population and that of ``pure'' AGN, i.e., they have values of both
the mass and the specific SFR that are intermediate between the two
classes. AGN, in addition, have a tail extending to very low values of
the specific SFR. Another, more physically motivated way of splitting
the AGN hosts is based on the strength (i.e., the accretion rate) of
the AGN itself, as indicated by its [OIII]$\lambda$5007\AA\
luminosity.  We use \citet{kauff_agn} dust-corrected
[OIII]$\lambda$5007\AA\ luminosities, and divide AGN hosts into strong
and weak AGN, using their demarcation of $L$[OIII] = $10^7
L_{\odot}$. We apply this division on all galaxies lying on the AGN
branch of the BPT diagram (we refer the reader to \citealt{kauff_agn}
for a visual presentation of where AGN of various strengths lie on the
BPT diagram). We show in Figure \ref{fig:ssfr_mass_agn2} the relation
between the specific SFR and mass for SF galaxies (thin solid
contours), strong (bold solid contours) and weak (dashed contours)
AGN. Apparently, strong AGN lie on the massive {\it continuation} of
the SF sequence, with SFRs as {\it high} as those of the massive
galaxies of the SF class (i.e., the offset from the dashed line is
similar). Higher masses with similar SFRs makes their {specific} SFRs
take more intermediate values. Weak AGN fall at lower star formation
rates relative to the strong AGN, with a tail toward very low star
formation rates that extends well into the domain of the quiescent
galaxies (see below).  \citet{kauff_agn}, using the optical
spectroscopy, found that strong AGN have younger populations than weak
AGN, which we qualitatively confirm, and provide a closer connection
between strong AGN and normal SF galaxies.

Here, we propose an evolutionary progression for massive galaxies:
from SF galaxies with no AGN, to strong AGN that are more massive yet
with similar SFRs, to weak AGN that have masses similar to strong AGN
but have lower (more suppressed) SFRs. Thus, the weak AGN would
represent a population with fading star formation. We emphasize that
while most optical and structural properties of the majority of AGN
would place them in the optical {\it red} sequence, we find that they
represent a smooth, massive extension of the star forming sequence
defined by optically {\it blue} galaxies. This picture is supplemented
by the results of \citet{kauff_galex}. They confirm that the majority
of blue $NUV-r$ galaxies with high velocity dispersion host AGN, and
suggest that strong AGN require extended gas-rich disks to provide a
reservoir of gas that is fueled into an AGN. Previously, without the
UV data, \citet{kauff_agn} suggested that many type 2 AGN are
post-starburst systems. Such a possibility, by itself, allows at least
two different evolutionary scenarios. One in which a normal
red-sequence galaxy experiences a burst (say, through a minor gas-rich
merger) which also fuels an AGN, and another, in which the AGN is
originally concurrent with normal SF. Our results, showing a smooth
sequence that begins with SF galaxies without AGN and extends at its
massive end to AGN hosts with different levels of star formation,
appear more compatible with the second scenario.

While neither \citet{kauff_galex} nor our study directly proves that
AGN are actually responsible for the subsequent quenching of the star
formation, they do provide more evidence for the connection between
the AGN and the fading star formation in massive galaxies. Such
picture is qualitatively compatible with recent theoretical work based
on $N$-body simulations of galaxy evolution and semi-analytic
treatment of feedback. Specifically, the ``radio''-mode AGN feedback,
proposed by \citet{croton}, demonstrates that the galaxies situated in
dark halos above some mass start to suppress the accretion of gas, and
at the same time suppress further star formation. These are called
``radio'' AGN because they can have black hole accretion rates well
below Eddington, as is the case in our sample. Also, this mechanism is
efficient in very massive galaxies. From the weighted plot of AGN
(lower middle panel of Figure \ref{fig:ssfr_mass_class}) we find that
the mass function of AGN peaks at around $4\times 10^{10} M_{\odot}$,
and has an abrupt turn-over at lower masses. This was noted by
\citet{kauff_agn} who also demonstrate that the low-mass cutoff is not
likely to be a selection effect; if a low-mass galaxy contained
anything but the very weakest AGN it would be classified as an AGN in
the BPT diagram. The connection that \galex\ has provided between the
SF galaxies and the AGN feedback is quite reassuring, and evidently
further work is required to complete this picture.

\subsection{Specific SFRs and stellar masses of galaxies with no 
\ha\ emission} \label{ssec:ssfr_noha}

Finally, we concentrate on galaxies for which no \ha\ emission was
detected in SDSS spectra. These galaxies have red {\it optical}
colors, which puts them in the ``red sequence'' category. These
galaxies are often referred to as ``old, red and dead'', because of
the presumably old populations and the lack of current SF. However, in
the UV to optical colors (Figure \ref{fig:cmd}) we saw that some of
these galaxies have relatively blue colors (which indicates that one
needs to be careful when using the term such as ``red sequence'',
since it apparently depends on the color used). In the right-hand
panels of Figure \ref{fig:ssfr_mass_class} we show specific SFRs of
galaxies with no \ha. In the unweighted plot they occupy a region of
the parameter space distinct from the star-forming and AGN galaxies,
with specific SFRs typically lower than for those two classes. The
volume-corrected plot still puts the bulk of them lower than the other
two classes, but the overlap with the low specific-SFR tail of the AGN
is more prominent. Also, the mass extent is now displayed accurately,
with the mass function turning over and flattening at lower masses.

If we expect most red-sequence galaxies to have no star formation at
all, how do we interpret the (specific) SFRs derived here? First, we
note that the specific SFRs (and also just the SFRs) are more poorly
constrained than for any other class (average formal error in specific
SFR is 0.63 dex). This should be interpreted in a way that the
observed SEDs are in many cases compatible with a very wide range of
SFRs, mostly very low ones. Let us note that our model SF histories do
not contain truncated models, but only exponentially declining models
(with added starbursts), which can never reach exact zero SFR. The
dynamic range of SFR in models is determined by the range of
exponential timescales, which in our case have a minimal e-folding
time of 1 Gyr. This produces a lower limit of $\log\, ({\rm
SFR}/L)\approx -14$, which is roughly equivalent to the lower limit in
the specific SFR shown in Figure \ref{fig:ssfr_mass_class}. While many
galaxies in this class are compatible with this lower limit, they do
not cluster close to it, because in the absence of stronger
constraints, higher SFRs will also be allowed (i.e., to firmly
constrain that the SFR is, say, smaller than $0.001\, M_{\odot} {\rm
yr^{-1}}$ one needs to measure a very low amount of UV light with a
very high precision). However, there are clear cases in which a galaxy
classified as having no \ha\ has a relatively well constrained
SFR. For some of them it is as high as $1\, M_\odot {\rm yr^{-1}}$
(crossing the dashed line). What is the morphology of galaxies with
high (specific) SFRs, yet no \ha\ emission? There are 837 galaxies
with $\log\, M_*>10$ and $\log\, ({\rm SFR}/M_*)>-11$. We inspect SDSS
images of these galaxies. Most appear as normal early-type galaxies
(with several obvious E+A types), which was already indicated in
\citet{rich} who find that some morphologically and spectroscopically
early-type galaxies show blue UV-to-optical colors. However, one
quarter shows some sign of structure or a disturbed light
profile. This fraction would very likely be higher in deeper images or
images with higher spatial resolution. One quarter of galaxies with
structure show traces of spiral arms. In most cases these spiral arms
are faint, and possibly represent cases where the star formation was
recently shut down (leading to the lack of emission lines), yet the
residual UV light and the spiral structure remain. Such galaxies have
been recognized in galaxy clusters as ``passive spirals''
\citep{anemic,passive}, and are believed to represent a transition
between a spiral and an S0 galaxy. In the inspected sample these
galaxies are found at all redshifts, indicating that the reason \ha\
is not detected is most likely not due to the aperture size being
smaller than the size of the galaxy from which the integral flux is
measured (especially for more nearby galaxies). Many of the galaxies
with structure exhibit rings---from very faint ones to bright blue
ones. In some of those cases SF could be occurring outside of the
spectroscopic fiber, leading to non-detection of \ha.

One can be concerned that our models attribute the UV light from old
populations (the UV ``upturn'') to young stars, thus implying an
artificially high star formation rate. \citet{bc03} models do include
post-AGB phases, but not the extreme blue horizontal branch, which is
believed (by no means unanimously) to be responsible for the UV upturn
in early-type galaxies \citep{brown}. However, \citet{donas} explore
nearby ``pure'' elliptical galaxies and find that both the UV and the
optical colors that they observe are fully reproduced in the models
that we use in this work, and that if the UV excess from the actual
star-formation is present, it is discernible from the combination of
UV and optical colors.  Therefore, our models do appear to account for
UV light from old populations and therefore some of the objects with
no \ha\ emission are quite likely to have had actual recent SF. This
conclusion is also reached in other studies based on \galex\ data
\citep{yi,kaviraj,schawinski}, where the blue UV-to-optical colors of
morphologically selected early-type galaxies are interpreted as signs
of recent star formation. It is quite conceivable that these galaxies
are in a post-starburst phase, so that the \ha\ emission is no longer
present, while the traces of star formation are still visible in the
UV. In evolutionary terms, as in the case of AGN, a question remains
whether this recent SF is the sign of a previously normal SF that is
now quenched, or is it the result of a small SF episode in an already
quiescent galaxy. In the latter case, the starburst could be due to a
minor merger with a gas-rich galaxy. Most likely both processes are
present, and the forthcoming work will address the relative importance
of each.

\subsection{Relationships between specific SFR and mass} \label{ssec:quan}

One would like to quantify the trends of the specific SFR trends with
respect to mass in order to facilitate easier comparison of our
optically-selected, local-universe sample to studies at other
redshifts, or samples selected in a different manner (keeping in mind
that a conversion may be required for a different choice of IMF or
$H_0$).

If one can identify ``pure'' star forming galaxies (with no AGN
contribution) using the BPT diagram, then the star forming sequence is
best described with a linear fit:

\begin{equation}
\log\, ({\rm SFR}/M_*) = -0.35(\log\, M_*-10)-9.83. 
\label{eqn:ssfr_mass_sf}
\end{equation}

\noindent In many instances one does not have all the requisite
spectral lines to make an exact classification, but instead relies on
color to separate star-forming galaxies. Adopting a color cut of
$NUV-r=4$, which passes in between the blue and the red sequence, and
taking galaxies on the blue side, we encompass all ``pure'' star
forming galaxies, but also many SF/AGN composites (see Figure
\ref{fig:cmd}. In that case, the star-forming sequence is better
described with a piecewise linear fit:

\begin{equation}
\log\, ({\rm SFR}/M_*) = \left\{\begin{array}{ll}
                       -0.17(\log\, M_*-10)-9.65, & \log\, M_*>9.4 \\
                       -0.53(\log\, M_*-10)-9.87, & \log\, M_*\leq 9.4.
\end{array} \right.
\label{eqn:ssfr_mass_blue}
\end{equation}

\noindent This relation differs from Equation \ref{eqn:ssfr_mass_sf}
mostly because the high-mass end is somewhat lower---a consequence of
including SF/AGN composites which have lower specific SFRs than pure
SF galaxies of the same mass. In both cases we have performed
least-square fitting to the {\it conditional} volume-corrected
composite PDFs (analogous to the upper panel of Figure
\ref{fig:ssfr_mass_cond_all}), binned into running modes in the
$8.1<\log\, M_*<11.7$ range for SF galaxies, and $8.0<\log\, M_*<11.6$
range for blue galaxies.

The trend of the specific SFR with respect to mass for the {\it
entire} sample is best observed in Figure
\ref{fig:ssfr_mass_cond_all}. We find that the parametrization having
the form similar to \citet{schechter} luminosity function provides an
excellent overall fit to this trend. Recently, \citet{feulner} have
used Schechter functions to describe the upper {\it envelopes} of
their SFR/$M_*$ vs.\ $M_*$ data points, but here we find that it
represents a good description of the {\it main} trend. We use the
Schechter function of the form:

\begin{equation}
{\rm SFR}/M_* = ({\rm SFR}/M_*)_0\, 10^{(\alpha+1)(\log\, M_*-\log\, M_0)}
\exp[-10^{(\log\, M_*-\log\, M_0)}],
\label{eqn:schechter1}
\end{equation}

\noindent where $({\rm SFR}/M_*)_0$, $M_0$, and $\alpha$ describe the
normalization, characteristic stellar mass, and the low-mass slope,
respectively. Taking modes of the conditional distribution in the
$8.0<\log\, M_*<11.9$ range, we find the following best-fitting
parameters:

\begin{equation}
\begin{array}{l}
({\rm SFR}/M_*)_0 = 5.96\times10^{-11}\,{\rm yr^{-1}}, \\
\log\, M_0 = 11.03, \\
\alpha = -1.35.
\end{array}
\label{eqn:schechter2}
\end{equation}

\noindent This fit, and the data points that were used to construct
it are shown in the lower panel of Figure
\ref{fig:ssfr_mass_cond_all}.

In the paradigm of ``downsizing'', at higher redshifts the SF was more
active in galaxies with larger mass. This should be reflected in the
relative flattening of the SF sequence at higher redshifts. While most
current studies observe an overall shift towards higher specific SFRs,
the slope does not appear to change much (e.g.,
\citealt{noeske,papovich}). Our quantification of the specific SFR
should enable more direct comparison with the local universe. In many
cases, simple overplotting of Equations \ref{eqn:schechter1} and
\ref{eqn:schechter2} to other sets of data would be indicative (unless
data are color-selected to include only the blue sequence, in which
case we recommend using Equation \ref{eqn:ssfr_mass_blue}).

\begin{figure}
\epsscale{1.2} 
\plotone{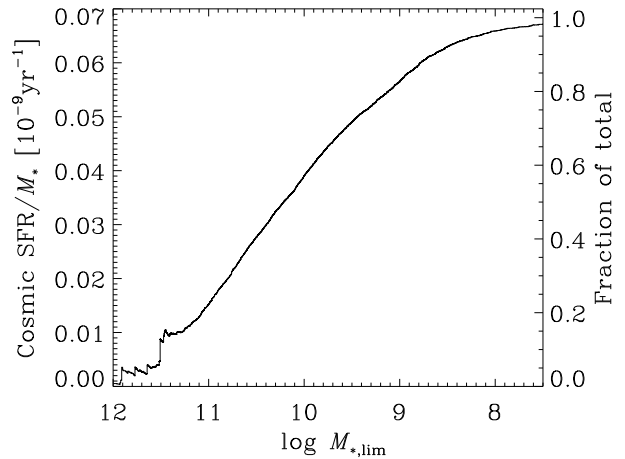}
\caption{Cosmic specific SFR as a function of the limiting mass. We
  define cosmic specific SFR as the the ratio of total SFR to total
  mass, each weighted by \vmax. The figure shows the effect of the
  limiting mass on the estimate of the cosmic specific SFR. SFRs have
  been corrected to $z=0.1$ by applying an evolution correction of the
  form $\propto (1+z)^{2.5}$.}
\label{fig:ave_ssfr}
\end{figure}

One often wishes to determine the average specific SFR of their
sample. The average specific SFR can be defined in several ways, but
the most used definition is one in which the total SFR in a
volume-limited sample is simply divided by the total stellar mass. We
call this the {\it cosmic} specific SFR:

\begin{equation}
{\rm SFR}/M_*({\rm Cosmic}) = \frac{\sum ({\rm ^{0.1}SFR}/V_{\rm max})}
{\sum ({M_*}/V_{\rm max})},
\end{equation}

\noindent where ${\rm ^{0.1}SFR}$ is the SFR corrected to $z=0.1$ (see
\S \ref{sec:sfrd}). From our sample we find ${\rm SFR}/M_*({\rm
Cosmic}) = 6.8\times 10^{-11}\, {\rm yr^{-1}}$, which is in very good
agreement with the equivalent measure in B04. How much is the estimate of
the cosmic specific SFR dependent on the limiting mass of the sample?
In Figure \ref{fig:ave_ssfr} we plot cosmic specific SFR as it would
have been determined had we had a mass limit (and assuming the same
magnitude limits). Right-hand $y$-axis shows a fraction of the
``actual'' (no mass limit) cosmic ${\rm SFR}/M_*$ at the given mass
limit.

Finally, we ask with what precision can one use the relations to
``predict'' specific SFR from stellar mass? For blue galaxies, we find
the average scatter around the mean sequence to be 0.5 dex. This
result is only slightly affected after taking out in quadrature the
formal measurement error of 0.2 dex. Therefore, the sequence has an
intrinsic width, and is likely to depend on other factors besides the
mass.  The width itself is a function of mass. It declines at the rate
of $-0.11$ dex$^{-1}$, i.e., the scatter of the specific SFR at $\log
M_* =8.0$ is 65\% larger than at $\log\, M_* =10.5$. This is an
indication that the star formation becomes less stochastic as the mass
increases, which is what one expects considering that massive galaxies
have a larger number of star-forming regions than dwarf galaxies.

\section{Global star formation density} \label{sec:sfrd}

In this section we use our SFRs to determine the global star formation
density at $z=0.1$, the mean redshift of our sample. In order to take
into account that the sample has a spread of redshifts, and therefore
probes a range of cosmic star formation history, in the remainder of
the section each galaxy's SFR estimate is ``corrected '' to the mean
redshift of the survey, where the uncertainties in the correction are
minimized. To make the correction, one assumes a cosmic star formation
density evolution of the form $(1+z)^{\beta}$, where B04 take
$\beta=3$. Therefore, the corrected SFR for a galaxy lying at redshift
$z$ is ${\rm ^{0.1}SFR} = [1.1/(1+z)]^{\beta}\, {\rm SFR}$.

Since we will be comparing our results to those obtained by B04, we
first describe their methodology and apply it to our data. B04 obtain
star formation density ($\rho_{\rm SFR}$), and its formal (random)
error, by Monte Carlo method applied to SFR PDFs in the following
way. First, 100 different bootstrap samples are constructed from the
original sample, using drawings with replacements. Bootstrap samples
have the same number of objects as the original sample. A random
number is then used to select a value from the cumulative PDFs for
each galaxy in the bootstrap sample. The random sampling of the PDF is
repeated 30 times. Each SFR PDF is weighted by $1/V_{\rm max}$ and
summed to get $\rho_{\rm SFR}$. From the ensemble of such samplings
the mode and the 16\% and 84\% confidence levels are determined for
the entire sample, and for each galaxy class. The results of this
exercise are presented in B04 Table 3. The random error of the total
$\rho_{\rm SFR}$ is found to be around 0.018 in units of $10^{-2}\,
h_{70}\, M_{\odot} {\rm yr}^{-1} {\rm Mpc}^{-3}$. For our sample, we
perform only the multiple probing of each galaxy's cumulative PDF,
without bootstrapping the sample. Performing 100 Monte Carlo
summations, we find a random error of the total $\rho_{\rm SFR}$ to be
0.017, in the same units. Therefore, the two studies have comparable
formal random errors, which is not surprising given the fact that the
formal errors of the SFRs are comparable. In any case, these random
errors are smaller than some systematic uncertainties that we will
explore later.

\begin{deluxetable}{lrrrrr}
\tablewidth{0pt}
\tablecaption{Comparison of SFR densities at $z=0.1$ in this study and
  in B04, derived using compatible methodology}
\tablehead{ Class & $\rho_{\rm SFR}$ & \% &
  $\rho_{\rm SFR}$ & \%  & fractional \\
 & This paper & of total & B04 & of total & change }

\startdata
All             &  1.953  &  100.0\%  &  1.980  &  100.0\%  &  0.99\\
SF              &  1.395  &   71.4\%  &  1.013  &   51.2\%  &  1.38\\
SF (low S/N)    &  0.296  &   15.2\%  &  0.455  &   23.0\%  &  0.65\\
Comp            &  0.147  &    7.5\%  &  0.207  &   10.5\%  &  0.71\\
AGN             &  0.076  &    3.9\%  &  0.084  &    4.2\%  &  0.90\\
No \ha          &  0.038  &    2.0\%  &  0.221  &   11.2\%  &  0.18\\
\enddata

\tablecomments{For this comparison we adopt $\beta=3$ SFR density
  evolution. SFR density is given in units of $10^{-2} h_{70}
  M_{\odot} {\rm yr}^{-1} {\rm Mpc}^{-3}$}
\label{tab:sfrd_comp}
\end{deluxetable}

How do the actual values of SFR density compare?  First, we notice that
the B04 method is essentially equivalent to the summation of the
medians of the PDFs since the average value of a random number that
probes the cumulative PDF is 0.5. In Table \ref{tab:sfrd_comp} we
compare the values obtained from the summation of the medians of our
SFR PDFs, to the average of the 16 and 84 percentiles of B04 values
(which should be close to the median). Our estimate includes the
correction for galaxies excluded from the sample because of the poor
SED fits (\S \ref{ssec:fit}). These corrections are 1.6--2.8\%,
depending on the class. Looking at the first row, we see that our
estimate of the total SFR density is very close to that of
B04. However, this is a mere coincidence when we look at how the
densities compare by class. Our SFR density for the SF class is 38\%
higher; however, for all other classes the estimate is lower---between
10\% less for AGN to 5.7 times less for galaxies without \ha. The
causes for these differences can be traced from the analysis of SFRs
in \S \ref{sec:sfr}. Consequently, the contribution of each class to
the total SFR density has changed as well. SF class now contributes
some 70\%, while it was around 50\% in B04. This comes at the expense
of other classes (except for the AGN), and most significantly from
much lower contribution of galaxies without \ha\ (2\%, vs.\ 11\% in
B04).

We next describe the calculation of our SFR density, which somewhat
differs from B04. Direct summation that was used to find SFR density
in B04 assumes that the losses from the incompleteness of the sample
are insignificant. B04 estimates them to be between 1 and 2
percent. We fit a Schechter fit to the number density of galaxies of
various classes and integrate the luminosity density down to our
completeness limit of $M_r=-14.75$, and to a limit which is 10
magnitudes fainter. We compare the two total luminosity densities and
find that 1.1\% of luminosity density is below our limit for SF (and
low-S/N SF) galaxies. For other classes, the loss is practically
zero. Therefore, for our SFR density estimate we perform the direct
summation by taking galaxies $M_r<-14.75$, and then correcting the SFR
density. This result differs insignificantly from one in which we
would sum all galaxies with no magnitude limit. Our SFR density
estimates are obtained by summing the {\it averages} of the SFR
estimates. We prefer summing averages over medians because it is
equivalent to summing an entire PDF. Averages produce a result that
is 5\% higher than the medians. Also, we adopt the exponent of the SFR
density evolution $\beta=2.5\pm 0.7$, determined by \citet{schim} from
\galex-VVDS measurements.

\begin{figure}
\epsscale{1.2}
\plotone{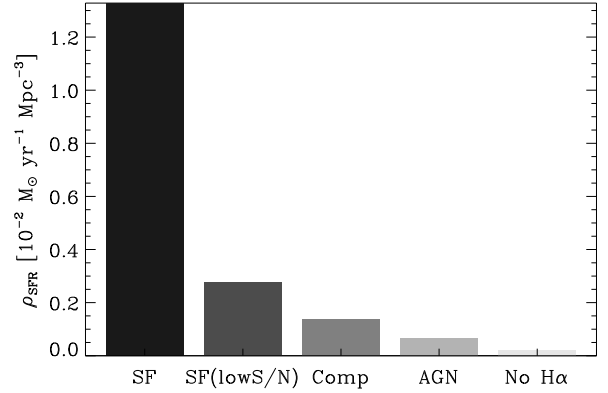}
\caption{Total star formation rate densities for different classes of
  galaxies. 88\% of total star formation occurs in ``normal'' SF
  galaxies (including low-S/N SF class), and another 11\% in galaxies
  that host an AGN. Up to 1\% occurs in galaxies that are quiescent
  based on the optical spectroscopy (no \ha\ class).Estimates are
  given for the mean redshift of the sample of $z=0.1$, by applying an
  evolution correction of the form $\propto (1+z)^{2.5}$.}
\label{fig:sfrd_class}
\end{figure}

\begin{deluxetable}{lrr}
\tablewidth{0pt}
\tablecaption{SFR densities at $z=0.1$ determined in this study}
\tablehead{ Class & $\rho_{\rm SFR}$ & \% of total }

\startdata
All             &  1.828  &  100.0\%  \\
SF              &  1.328  &   72.6\%  \\
SF (low S/N)    &  0.277  &   15.2\%  \\
Comp            &  0.136  &    7.5\%  \\
AGN             &  0.067  &    3.6\%  \\
No \ha          &  0.021  &    1.1\%  \\
\enddata

\tablecomments{We adopt $\beta=2.5$ SFR density evolution. SFR density is
  given in units of $10^{-2} h_{70} M_{\odot} {\rm yr}^{-1} {\rm
  Mpc}^{-3}$. Random error of SFR density for 'All' galaxies is 0.017 in
  same units. See text for the discussion of systematic errors.}
\label{tab:sfrd}
\end{deluxetable}

The results, and the breakdown by class, are presented in Table
\ref{tab:sfrd}, and shown in Figure \ref{fig:sfrd_class}. These are
our nominal SFR densities. Galaxies for which no classification was
possible (see \S \ref{ssec:bpt}) are included in the SF class. They
constitute 0.8\% of the total SFR densuty. Altogether, 88\% of star
formation occurs in ``normal'' SF galaxies, and another 11\% in
galaxies that host an AGN. Up to 1\% occurs in galaxies that are
quiescent based on the optical spectroscopy.

The uncertainty in the evolution exponent leads to the systematic
uncertainty in the total SFR density of 2\%. By far the largest
uncertainty stems from the systematic uncertainties in the
determination of the dust attenuation, and hence the SFR of SF
galaxies (\S \ref{ssec:comp_sf}). We estimate the degree of this error
by simply replacing the SFRs of SF galaxies by those obtained by
B04. This error is preferentially in the direction that the ``\ha''
estimate becomes 11\% higher for the SF class, leading to the total
density that is 8\% higher. We also take into account that our SFR
estimates are averaged over $10^8$ yr. Using $10^7$ yr instead, we
obtain a SFR density that is 1\% lower, which we will treat as the
systematic error. Finally, another small source of systematic error
comes from the error in the estimate of survey area of 0.3\%. Adding
up in quadrature random and systematic errors in the positive and in
the negative directions, we arrive at our total, dust-corrected SFR
density estimate at $z=0.1$:

\begin{equation}
\rho_{\rm SFR} = 1.828^{+0.148}_{-0.039} \times 10^{-2} h_{70}
M_{\odot} {\rm yr}^{-1} {\rm Mpc}^{-3},
\end{equation}

\noindent which assumes the Chabrier IMF. For a compilation of recent
measurements of the local $\rho_{\rm SFR}$ please refer to
\citet{hanish}. Our $\rho_{\rm SFR}$ agrees very well ($<1 \sigma$)
with four of the six estimates listed in their Table 1, but has
significantly smaller error bars. Note that except for B04, all other
estimates are based on samples with fewer than 300 galaxies, and many
apply fixed dust correction for all galaxies. We can find the average
FUV attenuation by ``de-correcting'' our SFRs by adding our FUV
attenuation estimates to each galaxy.  Such uncorrected SFR density is
smaller by a factor of 6.34. In other words, the average FUV
attenuation for the {\it volume-complete} sample is $A_{FUV}=2.01$
mag.

\begin{figure}
\epsscale{1.2}
\plotone{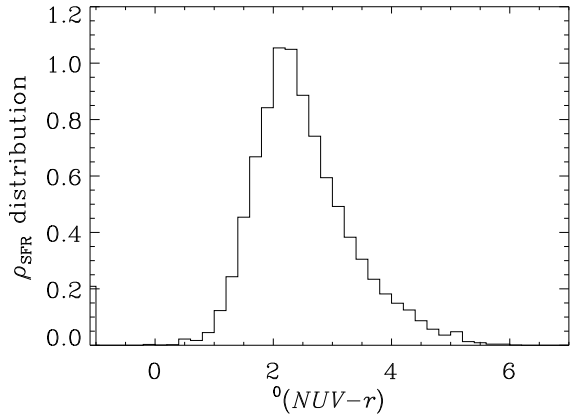}
\caption{Distribution of the star formation density (at $z=0.1$) as a
  function of color. We choose UV-to-optical color which shows a
  distinct bimodality in the number density of galaxies. The SFR
  density distribution is unimodal, and peaks sharply. There is a tail
  towards the red galaxies, which is in part due to the intrinsic dust
  reddening. A large fraction (94\%) of the star formation density
  arises from galaxies blueward of $NUV-r=4$---our nominal division
  between the blue and red galaxies. First bins includes all objects
  without NUV detection. Objects below our absolute magnitude
  completeness limit ($\sim 1\%$ of SFR density) are not included. The
  units of SFR density distribution are $10^{-2} h_{70} M_{\odot} {\rm
  yr}^{-1} {\rm Mpc^{-3} mag^{-1}}$.}
\label{fig:sfrd_nuv_r}
\end{figure}

We next explore the distribution of global SFR density across the
galaxies having different properties. In Figure \ref{fig:sfrd_nuv_r}
we show the distribution of SFR density against the rest-frame $NUV-r$
color. The distribution sharply peaks at $^{0}(NUV-r)=2.2$. While the
number density distribution is strikingly bimodal in this color, the
SFR distribution is very even. The blue edge is more abrupt then the
red side, which shows a tail extending to red-sequence colors, in part
due to the intrinsic dust reddening. We have previously used an
$NUV-r=4$ cut, which goes across the green valley, to conveniently
divide the blue and the red sequences. From Figure
\ref{fig:sfrd_nuv_r} we see that the large fraction (94\%) of star
formation occurs blue of this cut. The first bin sums SFR density in
galaxies lacking $>3\sigma$ measurement in NUV (2\% of total).

\begin{figure}
\epsscale{1.2}
\plotone{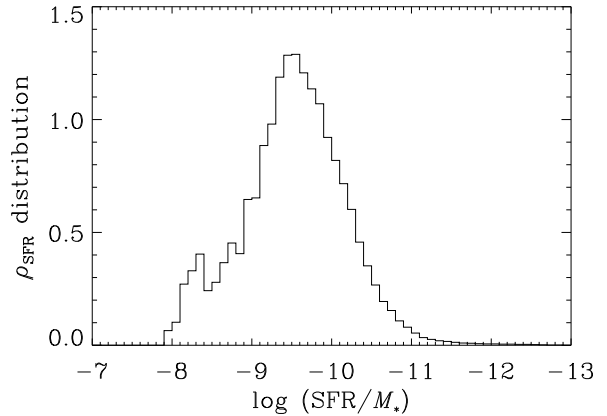}
\caption{Distribution of the star formation density (at $z=0.1$) in
  galaxies with different specific SFRs. Specific SFR is an indicator
  of a galaxy's star formation history. The distribution shows a
  ``bump'' at the high end due to the population of bursty, low-mass
  galaxies. There is a steep decline towards the galaxies with low
  specific SFR. The units of SFR density distribution are $10^{-2}
  h_{70} M_{\odot} {\rm yr}^{-1} {\rm Mpc^{-3} dex^{-1}}$.}
\label{fig:sfrd_ssfr}
\end{figure}

UV to optical color is easy to measure, but gives only a very rough
idea of a galaxy's SF history. Therefore, in Figure
\ref{fig:sfrd_ssfr} we plot the SF density distribution against the
specific SFR. The picture is somewhat different than from the
distribution against the color. First, the high specific-SFR end does
not fall as evenly as did the blue $NUV-r$ color end; instead, we see
an excess of galaxies with $\log\, ({\rm SFR}/M_*) \approx
-8.3$. These are the bursty, low-mass galaxies mentioned in \S
\ref{ssec:ssfr_sf}. The reason that $NUV-r$ color does not distinguish
this population is because it ``saturates'' for galaxies with $\log\,
({\rm SFR}/M_*) \approx -9$, so that all galaxies with higher specific
SFR will have similar $NUV-r$ colors. FUV (included in the specific
SFR) is needed to break the degeneracy. On the side of the low
specific SFR (equivalent to redder $NUV-r$ colors), the decline is
more steep than for the colors. The difference arises from the fact
that the specific SFR, unlike $NUV-r$ color, does take out the effects
of the intrinsic dust attenuation.

\begin{figure}
\epsscale{1.2} 
\plotone{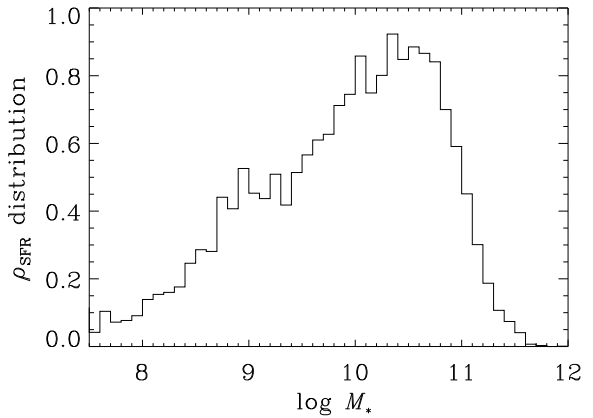}
\caption{Distribution of the star formation density (at $z=0.1$) in
  galaxies of different stellar mass. The fractional star formation
  density increases with galaxy's stellar mass and reaches a maximum
  at $\approx 4\times 10^{10} M_{\odot}$, which is close to the
  ``transitional'' mass of \citet{kauff}. Our sample is incomplete at
  $\log\, M_*<7.5$. The units of SFR density distribution are $10^{-2}
  h_{70} M_{\odot} {\rm yr}^{-1} {\rm Mpc^{-3} dex^{-1}}$.}
\label{fig:sfrd_mass}
\end{figure}

\begin{figure}
\epsscale{1.2} 
\plotone{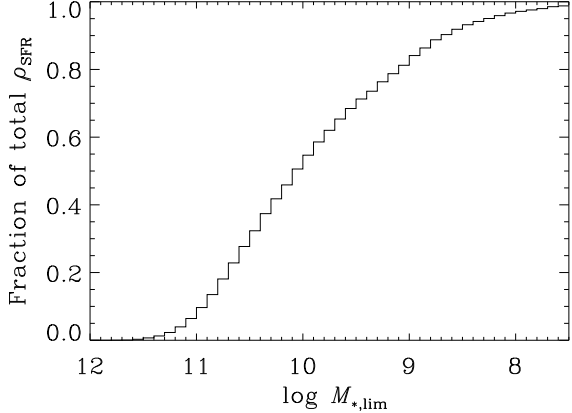}
\caption{Fraction of the total star formation density (at $z=0.1$) as
  a function of limiting stellar mass. The summing is performed from
  the high-mass end. Therefore, at each mass, the plot shows how much
  SF is missed by not including less massive galaxies. Our sample is
  incomplete at $\log\, M_*<7.5$, where we estimate 1\% of SF takes
  place.}
\label{fig:csfrd_mass}
\end{figure}

Finally, in Figure \ref{fig:sfrd_mass} we show the distribution of SFR
density across the galaxies with different stellar masses. The
fraction of SF occurring in galaxies with increasing mass rises almost
linearly. The peak of the SF today occurs in galaxies having a mass of
$\approx 4\times 10^{10} M_{\odot}$. This mass is close to the
``transitional'' mass of \citet{kauff}, which is effectively a mass at
which blue and red mass functions cross, i.e., the mass above which
the red galaxies become more numerous. In the paradigm of downsizing,
we would expect this distribution to move towards the higher masses at
higher redshifts. This, however, is often difficult to establish since
the completeness limits of many high-redshift surveys are
uncomfortably close to the peak of the distribution whose position one
tries to establish. In Figure \ref{fig:csfrd_mass} we show the
cumulative SFR density as a function of stellar mass, summing up from
the {\it highest} mass galaxies. This figure shows how much SF is
missed (in the local universe) by being limited to a given stellar
mass (and assuming the same magnitude limits). Thus, a limit of
$10^{11} M_{\odot}$ captures only 6\% of the total SFR density. In
order to reach 50\%, one needs to probe galaxies more massive than
$10^{10} M_{\odot}$. Today, 1/2 of the total SFR density comes from
galaxies in the mass range $9.3<\log\, M_*<10.6$.

\section{Conclusions}

We have considered the measurement of the star formation rate using a
sample of 50,000 optically selected galaxies with \galex\ and SDSS
photometry and SDSS redshifts. We have given a detailed description of
our approach to SED fitting, including the use of a large library of
population models that include a physically motivated treatment of
dust attenuation.  These give dust-corrected SFRs (``UV'' SFRs) of
galaxies in the local universe.  We show that there are no inherent
problems that would prevent UV from serving as a reliable SFR
indicator, despite past concerns over population synthesis and dust
attenuation models.  We recite some of the highlights from our study.

\begin{itemize}
\item 25 min imaging from \galex\ in two UV bands can be used to
derive SFRs with a formal accuracy of 0.2 dex for $r<18$ mag
star-forming galaxies with known redshifts.

\item For galaxies for which the dominant ionizing source are young
stars, (as determined using $3''$ fiber measurements and the BPT
diagram) the SFRs obtained through SED fitting (``UV'' SFRs) agree
exceptionally well with dust-corrected SFRs obtained from modeling of
the nebular lines---essentially \ha, corrected using a Balmer
decrement (``\ha'' SFRs), as long as the same dust model is applied to
both methods. The comparison presented here is at a scale that is two
to three orders of magnitude larger than the comparison of these SF
indicators done in the past.

\item The \citet{cf00} dust attenuation model that we apply to ``UV''
SFRs and that \citet{b04} apply to ``\ha'' SFRs succeeds in producing
UV and \ha\ dust attenuation estimates that lead to compatible SFRs.

\item Minor differences between the ``UV'' and ``\ha'' SFRs exist at
the extremes of galaxy stellar mass distributions (equivalent to the
extremes of the dust attenuations). They arise either from somewhat
weaker constraints from the SED-derived dust attenuation, or from the
mass-dependent deviations of the assumed $\propto \lambda^{-0.7}$
extinction law in the UV regime.

\item If the systematics in the attenuation estimates are accounted
for, the ``UV'' and ``\ha'' SFRs agree within 10\% across the entire
range of galaxy masses.

\item We calibrate the relations that allow SFRs to be determined
directly from K-corrected $FUV$ and $NUV$ magnitudes. They involve an
FUV attenuation estimate from the UV color (equivalent to the UV
slope) that we calibrated for normal star-forming galaxies, and find
it to be less steep than the \citet{meurer99} relation. To obtain SFRs
we find that one requires a conversion factor between the FUV
luminosity and SFR that is 30\% lower than \citet{kennicutt}
value. Such ``simple'' UV SFRs compare well with those obtained from
the full SED fitting in the 0.1--10 $M_{\odot} {\rm yr^{-1}}$ range,
i.e., for normal star-forming galaxies.

\item For some 50\% of the optically selected galaxies from SDSS, the
emission lines are either too weak (red sequence galaxies), or they
are contaminated by narrow-line AGN, which precludes obtaining SFRs
from \ha\ directly. In those cases \citet{b04} use calibrations
between SFR and optical colors from SF galaxies, and apply them to
other classes of galaxies. We find that such calibrations are not
reliable and lead to large overestimation of SFRs. UV, on the other
hand, provides more secure SFR estimates.

\item We present stellar-mass dependent star formation histories for
various classes of optically selected SDSS galaxies, from the
starbursting dwarf galaxies with specific SFRs (SFR/$M_*$) of $10^{-8}
{\rm yr^{-1}}$ to essentially dead, red-sequence galaxies with 6
orders of magnitude lower specific SFR. We also present a
volume-corrected version of these relationships which give a better
account of the SF at low masses. We quantify these relations in order
to facilitate comparison with similar results obtained at higher
redshifts.

\item Star forming galaxies with no AGN form a very well defined
linear star-forming sequence in specific SFR vs.\ mass plot, with the
intrinsic $1\, \sigma$ width of 0.5 dex in log~SFR/$M_*$. The sequence
is 65\% wider at the low mass end ($\sim 10^8 M_{\odot}$) than at high
masses, consistent with a more stochastic (bursty) SF history for
gas-rich dwarfs. The star-forming sequence terminates at $3\times
10^{11} M_{\odot}$.

\item Galaxies whose ionizing flux (as measured in $3''$ fiber) comes
primarily from young stars (star formation), those for which there is
a narrow-line AGN component (mostly Seyfert 2s and LINERs), and those
which are spectroscopically quiescent (no detection of \ha\ lines),
all occupy distinct positions in the specific SFR vs.\ stellar mass
diagram, with the AGN hosts lying in between SF galaxies with no AGN, and
the quiescent galaxies. Star forming galaxies with no AGN {\it do not}
occupy the intermediate regions of the specific SFR---those are
occupied exclusively by galaxies with AGN.

\item If we take all galaxies forming the AGN branch in the BPT
diagram (again from fiber measurements) and divide them by their AGN
strength (as indicated by [OIII]$\lambda$5007\AA\ luminosity), we find
that strong AGN represent the massive {\it continuation} of the SF
sequence defined by galaxies with no AGN.  Strong AGN have similar
SFRs as the most massive galaxies with no SF, but are more massive,
driving their specific SFRs to intermediate values. Weak AGN have
similar masses as strong AGN, but fall at lower star formation rates
relative to strong AGN, with a tail toward very low star formation
rates that extends well into the domain of the optically red galaxies
with no \ha.  While AGN appear to be a massive continuation of the
blue, star-forming sequence, their optical colors would place most of
them onto the {\it red} sequence. We suggest an evolutionary sequence
from massive star forming to quiescent galaxies via strong and weak
AGN that bridge them. This scenario supports a picture of AGN feedback
suppressing star formation in galaxies situated in very massive dark
matter halos, the so-called radio mode AGN \citep{croton}. Further
study is needed.

\item While mostly inactive, some galaxies with no detectable \ha\ in
their fiber measurements exhibit SFRs as high as $\sim 1 M_{\odot}
{\rm yr^{-1}}$. The optical colors of these galaxies are predominantly
red. We find that many (1/4) of such galaxies display disturbed light
profiles, sometimes in the form of fading spiral arms with no current
SF detectable in \ha. In many cases we see ring-like structures.  Some
of these are likely post-starburst systems. We find that up to 1\% of
UV SFR density comes from galaxies with no \ha\ lines.

\item We determine total SFR density in the local universe ($z=0.1$) of
$1.828^{+0.148}_{-0.039} \times 10^{-2} h_{70} M_{\odot} {\rm yr}^{-1}
{\rm Mpc}^{-3}$ (for Chabrier IMF), which is in good agreement with most
recent estimates, yet has a smaller error (which includes random and
most systematic errors).

\item Large fraction (95\%) of SF takes place in blue sequence
galaxies ($NUV-r<4$). SFR density increases in galaxies of increasing
mass, and peaks in galaxies with masses $\approx 4\times 10^{10}
M_{\odot}$---close to the ``transition'' mass of \citet{kauff} (i.e.,
the mass at which the mass functions of red (old) and blue galaxies
intersect). Galaxies in the mass range $9.3<\log\, M_*<10.6$ account for
1/2 of the total SFR density.
\end{itemize}

We demonstrate the ability of \galex\ UV observations of the galaxies
in the local universe to constrain SFRs of a statistically large
number of galaxies with a wide dynamic range of SF activity. Besides
characterizing the galaxies in the local universe, this study provides
a better reference point for high-redshift investigations. This and
the forthcoming studies based on \galex\ will continue to provide a
better understanding of the galaxy evolution, specifically the origin
of the galaxy bimodality, the inter-connection between the SF and AGN
activity, and the general star formation history of galaxies spanning
a range from gas-rich dwarfs to massive ellipticals.

\acknowledgments

We thank Janice C.\ Lee for insightful comments and helpful
discussions. We thank the referee for numerous valuable suggestions. 
We also thank Alessandro Boselli and Michael Blanton.
\galex\ ({\it Galaxy Evolution Explorer}) is a NASA Small Explorer,
launched in April 2003. We gratefully acknowledge NASA's support for
construction, operation, and science analysis for the GALEX mission,
developed in cooperation with the Centre National d'Etudes Spatiales
of France and the Korean Ministry of Science and Technology. Funding
for the Sloan Digital Sky Survey (SDSS) and SDSS-II has been provided
by the Alfred P. Sloan Foundation, the Participating Institutions, the
National Science Foundation, the U.S. Department of Energy, the
National Aeronautics and Space Administration, the Japanese
Monbukagakusho, and the Max Planck Society, and the Higher Education
Funding Council for England. This research has made use of NASA's
Astrophysics Data System.

{\it Facilities:} \facility{\galex, SDSS}


\begin{thebibliography}{}

\bibitem[Abazajian et al.(2004)]{sdssdr2} Abazajian, K., et 
al.\ 2004, \aj, 128, 502 
\bibitem[Adelman-McCarthy et al.(2006)]{sdssdr4} 
Adelman-McCarthy, J.~K., et al.\ 2006, \apjs, 162, 38 
\bibitem[Baldwin et al.(1981)]{bpt} Baldwin, J.~A., 
Phillips, M.~M., \& Terlevich, R.\ 1981, \pasp, 93, 5 
\bibitem[Baldry et al.(2004)]{baldry} Baldry, I.~K., 
Glazebrook, K., Brinkmann, J., Ivezi{\'c}, {\v Z}., Lupton, R.~H., Nichol, 
R.~C., \& Szalay, A.~S.\ 2004, \apj, 600, 681 
\bibitem[Bell \& Kennicutt(2001)]{bell_ken} Bell, E.~F., \& 
Kennicutt, R.~C., Jr.\ 2001, \apj, 548, 681 
\bibitem[Bell et al.(2004)]{bell} Bell, E.~F., et al.\ 2004, 
\apj, 608, 752 
\bibitem[Bertin \& Arnouts(1996)]{sextractor} Bertin, E., \& 
Arnouts, S.\ 1996, \aaps, 117, 393 
\bibitem[Blanton et al.(2005)]{vagc} Blanton, M.~R., et al.\
  2005, \aj, 129, 2562
\bibitem[Blanton \& Roweis(2007)]{blanton} Blanton, M.~R., \& 
Roweis, S.\ 2007, \aj, 133, 734 
\bibitem[Boissier et al.(2007)]{boissier} Boissier, S., et al.\ 
2007, \apjs\ Accepted, astro-ph/0609071 
\bibitem[Boselli et al.(2001)]{boselli} Boselli, A., Gavazzi, 
G., Donas, J., \& Scodeggio, M.\ 2001, \aj, 121, 753 
\bibitem[Brinchmann et al.(2004)]{b04} Brinchmann, J., 
Charlot, S., White, S.~D.~M., Tremonti, C., Kauffmann, G., Heckman, T., \& 
Brinkmann, J.\ 2004, \mnras, 351, 1151 
\bibitem[Brown et al.(2000)]{brown} Brown, T.~M., Bowers, 
C.~W., Kimble, R.~A., Sweigart, A.~V., \& Ferguson, H.~C.\ 2000, \apj, 532, 
308 
\bibitem[Bruzual \& Charlot(1993)]{bc93} Bruzual A., G., 
\& Charlot, S.\ 1993, \apj, 405, 538 
\bibitem[Bruzual \& Charlot(2003)]{bc03} Bruzual, G., \& 
Charlot, S.\ 2003, \mnras, 344, 1000 
\bibitem[Buat et al.(2002)]{buat02} Buat, V., Boselli, A., 
Gavazzi, G., \& Bonfanti, C.\ 2002, \aap, 383, 801 
\bibitem[Buat et al.(2005)]{buat05} Buat, V., et al.\ 2005, 
\apjl, 619, L51 
\bibitem[Calzetti et al.(1994)]{calzetti94} Calzetti, D., Kinney, 
A.~L., \& Storchi-Bergmann, T.\ 1994, \apj, 429, 582 
\bibitem[Chabrier(2003)]{chabrier} Chabrier, G.\ 2003, \pasp, 
115, 763 
\bibitem[Charlot \& Fall(2000)]{cf00} Charlot, S., \& Fall, 
S.~M.\ 2000, \apj, 539, 718 
\bibitem[Charlot \& Longhetti(2001)]{cl01} Charlot, S., \& 
Longhetti, M.\ 2001, \mnras, 323, 887 
\bibitem[Chester \& Roberts(1964)]{chester} Chester, C., \& 
Roberts, M.~S.\ 1964, \aj, 69, 635 
\bibitem[Cortese et al.(2006)]{cortese} Cortese, L., et al.\ 
2006, \apj, 637, 242 
\bibitem[Cram et al.(1998)]{cram} Cram, L., Hopkins, A., 
Mobasher, B., \& Rowan-Robinson, M.\ 1998, \apj, 507, 155 
\bibitem[MacKay (2003)]{mackay} MacKay, D.~J.~C., 2003, Information Theory, 
Inference, and Learning Algorithms, Cambridge University Press
Mobasher, B., \& Rowan-Robinson, M.\ 1998, \apj, 507, 155 
\bibitem[Croton et al.(2006)]{croton} Croton, D.~J., et al.\ 
2006, \mnras, 365, 11 
\bibitem[David et al.(1992)]{xray} David, L.~P., Jones, C., 
\& Forman, W.\ 1992, \apj, 388, 82 
\bibitem[de Jong et al.(1985)]{dejong} de Jong, T., Klein, U., 
Wielebinski, R., \& Wunderlich, E.\ 1985, \aap, 147, L6 
\bibitem[Donas et al.(2007)]{donas} Donas, J., et al.\ 
2007,  \apjs\ Accepted, astro-ph/0608594
\bibitem[Faber et al.(2005)]{sandy} Faber, S.~M., et al.\ 
2005, \apj\ Submitted, astro-ph/0506044 
\bibitem[Ferland(1996)]{ferland96} Ferland, G.~J.\ 1996, 
University of Kentucky Internal Report
\bibitem[Feulner et al.(2006)]{feulner} Feulner, G., Hopp, U., 
\& Botzler, C.~S.\ 2006, \aap, 451, L13 
\bibitem[Gavazzi \& Scodeggio(1996)]{gavazzi} Gavazzi, G., \& 
Scodeggio, M.\ 1996, \aap, 312, L29 
\bibitem[Gavazzi et al.(1996)]{gavazzi2} Gavazzi, G., Pierini, 
D., \& Boselli, A.\ 1996, \aap, 312, 397 
\bibitem[Gil de Paz et al.(2007)]{gildepaz} Gil de Paz, A., et 
al.\ 2007, \apjs\ Accepted, astro-ph/0606440
\bibitem[Hanish et al.(2006)]{hanish} Hanish, D.~J., et al.\ 
2006, \apj, 649, 150 
\bibitem[Hirashita et al.(2003)]{hirashita03} Hirashita, H., Buat, 
V., \& Inoue, A.~K.\ 2003, \aap, 410, 83 
\bibitem[Holmberg(1965)]{holmberg} Holmberg, E.\ 1965, Arkiv for 
Astronomi, 3, 387 
\bibitem[Hopkins et al.(2001)]{hopkins} Hopkins, A.~M., 
Connolly, A.~J., Haarsma, D.~B., \& Cram, L.~E.\ 2001, \aj, 122, 288 
\bibitem[Hubble(1926)]{hubble1} Hubble, E.~P.\ 1926, \apj, 64, 
321 
\bibitem[Hubble(1936)]{hubble2} Hubble, E.~P.\ 1936, Yale 
University Press.
\bibitem[Iglesias-P{\'a}ramo et al.(2004)]{paramo04} 
Iglesias-P{\'a}ramo, J., Boselli, A., Gavazzi, G., \& Zaccardo, A.\ 2004, 
\aap, 421, 887 
\bibitem[Johnson et al.(2007)]{ben} Johnson, B.~D., et 
al.\ 2007, \apjs\ Submitted
\bibitem[Kauffmann et al.(2003a)]{k03} Kauffmann, G., et 
al.\ 2003a, \mnras, 341, 33 
\bibitem[Kauffmann et al.(2003b)]{kauff} Kauffmann, G., et 
al.\ 2003b, \mnras, 341, 54 
\bibitem[Kauffmann et al.(2003c)]{kauff_agn} Kauffmann, G., et 
al.\ 2003c, \mnras, 346, 1055 
\bibitem[Kauffmann et al.(2007)]{kauff_galex} Kauffmann, G., et 
al.\ 2007, \apjs\ Accepted, astro-ph/0609436
\bibitem[Kaviraj et al.(2007)]{kaviraj} Kaviraj, S., et al.\ 
2007, \apjs\ Accepted, astro-ph/0601029 
\bibitem[Kennicutt(1983)]{kennicutt_radio} Kennicutt, R.\ 1983, \aap, 
120, 219 
\bibitem[Kennicutt(1998)]{kennicutt} Kennicutt, R.~C., Jr.\ 1998, 
\araa, 36, 189 
\bibitem[Kilgard et al.(2002)]{xray2} Kilgard, R.~E., Kaaret, 
P., Krauss, M.~I., Prestwich, A.~H., Raley, M.~T., \& Zezas, A.\ 2002, 
\apj, 573, 138 
\bibitem[Kroupa(2001)]{kroupa01} Kroupa, P.\ 2001, \mnras, 322, 
231 
\bibitem[Larson \& Tinsley(1978)]{lar_tin} Larson, R.~B., \& 
Tinsley, B.~M.\ 1978, \apj, 219, 46 
\bibitem[Leonard \& Hsu (2003)]{bayes} Leonard, T., \& Hsu, J.~S.~J.\ 2003, 
Bayesian Methods : An Analysis for Statisticians and 
Interdisciplinary Researchers, Cambridge University Press
\bibitem[Madau et al.(1996)]{madau} Madau, P., Ferguson, 
H.~C., Dickinson, M.~E., Giavalisco, M., Steidel, C.~C., \& Fruchter, A.\ 
1996, \mnras, 283, 1388 
\bibitem[Martin et al.(2005)]{chris_galex} Martin, D.~C., et al.\ 
2005, \apjl, 619, L1 
\bibitem[Martin et al.(2007)]{chris2} Martin, D.~C., et al.\ 
2007,  \apjs\ Accepted, astro-ph/0703281
\bibitem[Meurer et al.(1999)]{meurer99} Meurer, G.~R., Heckman, 
T.~M., \& Calzetti, D.\ 1999, \apj, 521, 64 
\bibitem[Milliard et al.(1992)]{milliard} Milliard, B., Donas, 
J., Laget, M., Armand, C., \& Vuillemin, A.\ 1992, \aap, 257, 24 
\bibitem[Morrissey et al.(2005)]{morrissey} Morrissey, P., et 
al.\ 2005, \apjl, 619, L7 
\bibitem[Morrissey et al.(2007)]{morrissey2} Morrissey, P., et 
al.\ 2007, \apjs\ Submitted
\bibitem[Noeske et al.(2007)]{noeske} Noeske, K.~G., et al.\ 2007,
  \apjl\ Accepted, astro-ph/0701924
\bibitem[Panuzzo et al.(2007)]{panuzzo} Panuzzo, P., Granato, 
G.~L., Buat, V., Inoue, A.~K., Silva, L., Iglesias-Paramo, J., \& Bressan, 
A.\ 2007, \mnras, 375, 640 
\bibitem[Papovich et al.(2006)]{papovich} Papovich, C., et al.\ 
2006, \apj, 640, 92 
\bibitem[Poggianti \& Barbaro(1997)]{poggianti} Poggianti, B.~M., 
\& Barbaro, G.\ 1997, \aap, 325, 1025 
\bibitem[Poggianti et al.(1999)]{passive} Poggianti, B.~M., 
Smail, I., Dressler, A., Couch, W.~J., Barger, A.~J., Butcher, H., Ellis, 
R.~S., \& Oemler, A.~J.\ 1999, \apj, 518, 576 
\bibitem[Rich et al.(2005)]{rich} Rich, R.~M., et al.\ 2005, 
\apjl, 619, L107 
\bibitem[Rosa-Gonz{\'a}lez et al.(2002)]{rosa-gonzales} 
Rosa-Gonz{\'a}lez, D., Terlevich, E., \& Terlevich, R.\ 2002, \mnras, 332, 
283 
\bibitem[Roussel et al.(2001)]{roussel} Roussel, H., Sauvage, 
M., Vigroux, L., \& Bosma, A.\ 2001, \aap, 372, 427 
\bibitem[Salim et al.(2005)]{salim} Salim, S., et al.\ 2005, 
\apjl, 619, L39 
\bibitem[Salpeter(1955)]{salpeter} Salpeter, E.~E.\ 1955, \apj, 
121, 161 
\bibitem[Sandage(1972)]{sandage} Sandage, A.\ 1972, \apj, 176, 
21 
\bibitem[Schawinski et al.(2007)]{schawinski} Schawinski, K., et 
al.\ 2007, \apjs\ Accepted, astro-ph/0601036 
\bibitem[Schechter(1976)]{schechter} Schechter, P.\ 1976, \apj, 
203, 297 
\bibitem[Schiminovich et al.(2005)]{schim} Schiminovich, D., 
et al.\ 2005, \apjl, 619, L47 
\bibitem[Scranton et al.(2005)]{scranton} Scranton, R., 
Connolly, A.~J., Szalay, A.~S., Lupton, R.~H., Johnston, D., Budavari, T., 
Brinkman, J., \& Fukugita, M.\ 2005, \aj\ submitted, 
astro-ph/0508564 
\bibitem[Searle et al.(1973)]{searle} Searle, L., Sargent, 
W.~L.~W., \& Bagnuolo, W.~G.\ 1973, \apj, 179, 427 
\bibitem[Seibert et al.(2005)]{seibert} Seibert, M., et al.\ 
2005, \apjl, 619, L55
\bibitem[Skrutskie et al.(2006)]{2mass} Skrutskie, M.~F., et 
al.\ 2006, \aj, 131, 1163 
\bibitem[Springel et al.(2005)]{springel} Springel, V., Di 
Matteo, T., \& Hernquist, L.\ 2005, \apjl, 620, L79 
\bibitem[Strauss et al.(2002)]{strauss} Strauss, M.~A., et al.\ 
2002, \aj, 124, 1810 
\bibitem[Sullivan et al.(2000)]{sullivan1} Sullivan, M., Treyer, 
M.~A., Ellis, R.~S., Bridges, T.~J., Milliard, B., \& Donas, J.\ 2000, 
\mnras, 312, 442 
\bibitem[Sullivan et al.(2001)]{sullivan2} Sullivan, M., 
Mobasher, B., Chan, B., Cram, L., Ellis, R., Treyer, M., \& Hopkins, A.\ 
2001, \apj, 558, 72 
\bibitem[Trammell et al.(2007)]{trammell} Trammell, G.~B., 
Vanden Berk, D.~E., Schneider, D.~P., Richards, G.~T., Hall, P.~B., 
Anderson, S.~F., \& Brinkmann, J.\ 2007, \aj, 133, 1780 
\bibitem[van den Bergh(1976)]{anemic} van den Bergh, S.\ 1976, 
\apj, 206, 883 
\bibitem[Wang \& Heckman(1996)]{wang} Wang, B., \& Heckman, 
T.~M.\ 1996, \apj, 457, 645 
\bibitem[Wyder et al.(2007)]{wyder} Wyder, T.~K., et 
al.\ 2007, \apjs\ Submitted
\bibitem[Yi et al.(2005)]{yi} Yi, S.~K., et al.\ 2005, 
\apjl, 619, L111

\end{thebibliography}
\end{document}